\documentclass[onecolumn]{aa}
\usepackage{graphicx,amsmath}
\usepackage{epsfig}

\newcommand{\bea}{\begin{eqnarray}}
\newcommand{\eea}{\end{eqnarray}}
\newcommand{\be}{\begin{equation}}
\newcommand{\ee}{\end{equation}}
\newcommand{\ben}{\begin{enumerate}}
\newcommand{\een}{\end{enumerate}}
\newcommand{\bi}{\begin{itemize}}
\newcommand{\ei}{\end{itemize}}
\newcommand{\bmi}[1]{\begin{minipage}{#1 cm}}
\newcommand{\emi}{\end{minipage}}

\newcommand{\rund}[1]{\left(#1\right)}
\newcommand{\vc}[1]{\mbox{\boldmath $#1$}}
\renewcommand{\d}{{\rm d}}
\newcommand{\eck}[1]{\left[ #1 \right]}
\newcommand{\ave}[1]{\left\langle #1 \right\rangle}
\newcommand{\wave}[1]{\left\{ #1 \right\}}

\newcommand{\abs}[1]{\left| #1 \right|}

\def\llabel#1{\label{sc:#1}}
\def\elabel#1{\label{eq:#1}}

{\catcode`\@=11
\gdef\SchlangeUnter#1#2{\lower2pt\vbox{\baselineskip 0pt \lineskip0pt
  \ialign{$\m@th#1\hfil##\hfil$\crcr#2\crcr\sim\crcr}}}
}
\def\gtrsim{\mathrel{\mathpalette\SchlangeUnter>}}

\begin{document}

\title{Constrained correlation functions}

\author{Peter Schneider\inst{1} and Jan Hartlap\inst{1}	}
\offprints{P. Schneider}

\institute{$^1$Argelander-Institut f\"ur Astronomie, Universit\"at Bonn,
				Auf dem H\"ugel 71, D-53121 Bonn, Germany\\
			\email{peter, hartlap@astro.uni-bonn.de}}
\titlerunning{Constrained correlation functions}
\date{Received ; accepted }

\abstract{

Measurements of correlation functions and their comparison with
theoretical models are often employed in natural sciences,
including astrophysics and cosmology, to determine best-fitting
model parameters and their confidence regions. Due to a lack of
better descriptions, the likelihood function of the correlation
function is often assumed to be a multi-variate Gaussian.

Using different methods, we show that correlation functions have
to satisfy contraint relations, owing to the non-negativity of
the power spectrum of the underlying random
process. Specifically, for any statistically homogeneous and (for
more than one spatial dimension) isotropic random field with
correlation function $\xi(x)$, we derive inequalities for the
correlation coefficients $r_n\equiv \xi(n x)/\xi(0)$ (for integer
$n$) of the form $r_{n{\rm l}}\le r_n\le r_{n{\rm u}}$, where the
lower and upper bounds on $r_n$ depend on the $r_j$, with
$j<n$, or more explicitly
\[
\Xi_{n-}\wave{\xi(0),\xi(x), \xi(2x),\dots,\xi([n-1]x)}\le \xi(n x)
\le \Xi_{n+}\wave{\xi(0),\xi(x), \xi(2x),\dots,\xi([n-1]x)} \;.
\]
Explicit expressions for the bounds are obtained for
arbitrary $n$. We show that these constraint equations very
significantly limit the set of possible correlation
functions. For one particular example of a fiducial cosmic shear
survey, we show that the Gaussian likelihood ellipsoid has a
significant spill-over into the region of correlation functions
forbidden by the aforementioned constraints, rendering
the resulting best-fitting model parameters and their error
region questionable, and indicating the need for a better description
of the likelihood function.

We conduct some simple numerical experiments which explicitly
demonstrate the failure of a Gaussian description for the
likelihood of $\xi$. Instead, the shape of the likelihood
function of the correlation coefficients appears to follow
approximately that of the shape of the bounds on the $r_n$, even
if the Gaussian ellipsoid lies well within the allowed region.
Therefore, we define a non-linear and coupled transformation of
the $r_n$, based on these bounds. Some numerical experiments then
indicate that a Gaussian is a much better description of the
likelihood in these transformed variables than of the original
correlation coefficients -- in particular, the full probability
distribution then lies explicitly in the allowed region.

For more than one spatial dimension of the random field, the explicit
expressions of the bounds on the $r_n$ are not optimal. We outline a
geometrical method how tighter bounds may be obtained in principle. We
illustrate this method for a few simple cases; a more general
treatment awaits future work.

}

\maketitle

\section{Introduction}
One of the standard ways to obtain constraints on model parameters of
a stochastic process is the determination of its two-point correlation
function $\xi(\vc x)$ from observational data, where $\vc x$ is the
separation vector between pairs of points.
This observed correlation function is then compared with the
corresponding correlation function $\xi(\vc x;p)$ from a model,
where $p$ denotes the model parameter(s). A commonly used method for this
comparison is the consideration of the likelihood function ${\cal
  L}(\xi|p)$, which yields the probability for observing the correlation
function $\xi(\vc x)$ for a given set of parameters $p$. It is common
(see Seljak \& Bertschinger 1993 for an application to microwave
background anisotropies, Fu et al.\ 2008 for a cosmic shear analysis,
or Okumura et al.\ 2008 for an application to the spatial
correlation function of galaxies)
to approximate this likelihood by a Gaussian,
\be
{\cal L}(\{\xi(x_i)\}|p)\propto \exp\eck{- {1\over 2} \sum_{i,j=1}^N
\eck{\xi(x_i)-\xi(x_i;p)}\,{\rm
  \tens{Cov}}^{-1}_{ij}\eck{\xi(x_j)-\xi(x_j;p)}}\;,
\elabel{LLikeli}
\ee
where it has been assumed that the random field is homogeneous and
isotropic, so that the correlation function depends only on the
absolute value of the separation vector. Furthermore, it has been
assumed that the correlation function is obtained at discrete points
$x_i$; for an actual measurement, one usually has to bin the
separation of pairs of points, in which case $x_i$ is the central
value of the bin. In (\ref{eq:LLikeli}), $\tens{Cov}$ is the
covariance matrix of the correlation function between any pair of
separations $x_i$, $x_j$.

In a recent paper (Hartlap et al.\ 2009) we have investigated the
likelihood function for the cosmic shear correlation function and
found that it deviates significantly from a Gaussian. This study
relied on numerical ray-tracing simulations through the density field
obtained from N-body simulations of the large-scale structure in the
Universe.

In this paper, we will show that the likelihood function of the
correlation function cannot be a Gaussian.  In particular, we show
that any correlation function obeys strict constraints, which can be
expressed as
\be
\Xi_{n-}\wave{\xi(0),\xi(x), \xi(2x),\dots,\xi([n-1]x)}\le \xi(n x)
\le \Xi_{n+}\wave{\xi(0),\xi(x), \xi(2x),\dots,\xi([n-1]x)}
\elabel{xiBounds}
\ee
for arbitrary $x$ and integer $n$; these constraints can be derived by
several different methods.  With one of these methods, one can derive
explicit equations for the upper and lower bounds in
(\ref{eq:xiBounds}) for arbitrary values of $n$.  The basic reason for
the occurrence of such constraints is the non-negativity of the power
spectrum, or equivalently, the fact that covariance matrices of the
values of a random fields at different positions are positive
(semi-)definite.

The outline of the paper is as follows: In Sect.\ts\ref{sc:2}, we
obtain bounds on the correlation function using the Cauchy--Schwarz
inequality, as well as making use of the positive definiteness of the
covariance matrix of random fields. It turns out that the latter
method gives tighter constraints on $\xi$; in fact, these constraints
are optimal for one-dimensional random fields. We show in
Sect.\ts\ref{sc:3} that these bounds significantly constrain the set
of functions which can possibly be correlation function. Whereas the
bounds obtained in Sect.\ts\ref{sc:2} are valid for any dimension of
the random field, they are not optimal in more than one dimension; we
consider generalizations to higher-order random fields, and to
arbitrary combinations of separations $x_i$ in Sect.\ts\ref{sc:4}. In
Sect.\ts\ref{sc:5} we introduce a non-linear coupled transformation of
the correlation coefficients based on the bounds; for the case of a
one-dimensional field, all combinations of values in these transformed
quantities correspond to allowed correlation functions (meaning that
one can find a non-negative power spectrum yielding the corresponding
correlations). Hence, a Gaussian probability distribution of these
transformed variables appears to be more realistic than one for the
correlation function itself. This expectation is verified with some
numerical experiments which are described in
Sect.\ts\ref{sc:6}. Furthermore, we show that for a fiducial
cosmological survey the Gaussian likelihood for the correlation
function can significantly overlap with the region forbidden by
(\ref{eq:xiBounds}), depending on survey size and number of
separations at which the correlation function is measured. We conclude
with a discussion and an outlook to future work and open questions.

\section{\llabel{2}Upper and lower bounds on correlation functions}
Consider an $n$-dimensional homogeneous and isotropic random field
$g(\vc x)$, with vanishing expectation value $\ave{g(\vc x)}=0$,
and with power spectrum $P(|\vec k|)$ and correlation function
\be
\xi(\vc x)=\int{\d^n k\over (2\pi)^n } \;P(|\vc k|)\,\exp\rund{-{\rm i}\vc
  k\cdot \vc x}\;,
\elabel{xiofP}
\ee
which depends only on the absolute value of $\vc x$, due to the
assumed isotropy. This relation immediately shows that
\be
-\xi(0)\le \xi(x)\le \xi(0) \;,
\elabel{constr1}
\ee
owing to $P(k)\ge 0$ and $\abs{\exp\rund{-{\rm i}\vc
  k\cdot \vc x}}\le 1$. However, the lower bound in (\ref{eq:constr1})
is not an optimal one for more than one dimension. In two
dimensions, the integral over the polar angle can be carried out,
yielding
\be
\xi_{\rm 2-D}(x)=\int_0^\infty {\d k\; k\over 2\pi}
\,P(k)\,{\rm J}_0(kx)\;,
\elabel{xiofP-2D}
\ee
where ${\rm J}_0$ is the Bessel function of the first kind of zero
order. Since ${\rm J}_0(x)$ has an absolute minimum at $x\approx 3.83$
with ${\rm J}_{0,{\rm min}}\approx -0.4028$ (see also Abrahamsen
1997), the non-negativity of $P(k)$ implies that $\xi_{\rm 2-D}(x)\ge
{\rm J}_{0,{\rm min}}\,\xi_{\rm 2-D}(0)$. Similarly, in three
dimensions one has
\be
\xi_{\rm 3-D}(x)=\int_0^\infty {\d k\; k^2\over 2\pi^2}\,
P(k)\,{\rm j}_0(kx)\;,
\elabel{xiofP-3D}
\ee
where ${\rm j}_0(x)=\sin x/x$ is the spherical Bessel function of zero
order. Since ${\rm j}_0(x)$ has an absolute minimum at $x\approx
4.493$ of ${\rm j}_{0{\rm min}}\approx -0.2172$, the non-negativity of
  $P(k)$ implies that $\xi_{\rm 3-D}(x)\ge {\rm j}_{0,{\rm
    min}}\,\xi_{\rm 3-D}(0)$.

In the following, we will concentrate mainly on the one-dimensional
case and write
\be
\xi(x)=\int_0^\infty \d k\;P_0(k)\,\cos(x k) \;.
\elabel{xiofP0}
\ee
However, higher dimensions of the random field are included in
all what follows, since by specifying $\vc x=(x,0,\dots,0)$ in
(\ref{eq:xiofP}), we find
\be
\xi(x)=\int{\d^n k\over (2\pi)^n } \;P\rund{\vc k}
\,\exp\rund{-{\rm i}
  k_1 \vc x}
=\int_0^\infty\d k_1\;\cos(k_1 x)\;
{2\over (2\pi)^n}\int\d k_2\dots\d
k_n\,P\rund{k_1,k_2,\dots,k_n} \;,
\elabel{mdto1d}
\ee
which thus takes the same form as (\ref{eq:xiofP0}). Thus, the
$n$-dimensional case can be included in the same formalism as the
one-dimensional case; note that for this argument, the random field is
not restricted to be isotropic. However, as we shall discuss later, the
resulting inequalities will not be optimal for isotropic fields of higher
dimension.

In the foregoing equations, $P(k)$ can correspond either to the power
spectrum of the underlying random process, or the sum of the
underlying process and statistical noise. Furthermore, the power
spectrum can also be the square of the Fourier transform of the
realization of a random process in a finite sample volume. In all
these cases, the non-negativity of $P(k)$ applies, and the constraints
on the corresponding correlation function derived below must hold. 

\subsection{Constraints from the Cauchy--Schwarz inequality}
Making use of the Cauchy--Schwarz inequality,
\be
\eck{\int_0^\infty \d k\; f(k)\,h(k)}^2
\le \int_0^\infty \d k\; f^2(k)\;\int_0^\infty \d k\; h^2(k)\;,
\ee
we obtain by setting $f(k)=\sqrt{P_0(k)}$ and
$h(k)=\sqrt{P_0(k)}\,\cos(x k)$ that\footnote{Note that this choice is
  possible because the power spectrum is non-negative!}
\be
\xi^2(x)\le \xi(0)\,\int_0^\infty \d k\;P_0(k)\,\cos^2(x k)
=\xi(0)\,\int_0^\infty \d k\;P_0(k)\,{1+\cos(2 x k)\over 2}
={\xi(0)\over 2}\eck{\xi(0)+\xi(2 x)}\;,
\ee
where we made use of the identity $\cos^2
a=\eck{1+\cos(2a)}/2$. Together with (\ref{eq:constr1}) we therefore
obtain the constraint equation
\be
-\xi(0)+{2 \xi^2(x)\over \xi(0)}\le \xi(2 x)
\le \xi(0)\;.
\elabel{constr2}
\ee
The interpretation of this constraint can be better seen in terms of
the correlation coefficient $r_n\equiv \xi(n x)/\xi(0)$, which is
defined for an arbitrary $x$. Then, (\ref{eq:constr2}) reads
\be
-1+2 r_1^2\le r_2\le 1\;.
\elabel{constr2r}
\ee
This result can be interpreted as follows: 
If two points separated by $x$ are strongly correlated, $1-r_1\ll 1$,
then the value of the field at a position $2 x$ must equally be
correlated with that at $x$, which implies that also the correlation
between the point $2 x$ and the origin must be large. If the field at
$x$ is strongly anticorrelated with that at the origin, $1+r_1\ll 1$,
than the field at $2x$ must be similarly anticorrelated with that at
$x$, implying a strong correlation between the point $2x$ and the
origin. The smaller $|r_1|$, the weaker is the constraint
(\ref{eq:constr2r}).

Making use of the identity
\[
\eck{\cos a+\cos(2a)}^2=\eck{1+\cos a}\,\eck{1+\cos (3a)}\;,
\]
and applying the Cauchy--Schwarz inequality with $f(k)=\sqrt{P_0(k)}
\sqrt{1+\cos(x k)}$ and $h(k)=\sqrt{P_0(k)}
\sqrt{1+\cos(3 x k)}$, we find that
\be
\eck{\xi(x)+\xi(2 x)}^2\le \eck{\xi(0)+\xi(x)}\,\eck{\xi(0)+\xi(3
  x)}\;.
\ee
A second inequality is obtained by using in a similar way the identity
\[
\eck{\cos a-\cos(2a)}^2=\eck{1-\cos a}\,\eck{1-\cos (3a)}\;.
\]
Both of these inequalities are summarized in terms of the correlation
coefficient as
\be
-1+{\rund{r_1+r_2}^2\over (1+r_1)} \le r_3\le
1-{\rund{r_1-r_2}^2\over (1-r_1)} \;.
\elabel{constr3r}
\ee
Further inequalities involving $\xi(m x)$, with $m\ge 4$ being an
integer, can be derived in this way. Making use of the relations
\[
[\cos a+\cos(n-1)a]^2=[1+\cos(n a)]\,[1+\cos(n-2)a]\; ; \;\;
[\cos a-\cos(n-1)a]^2=[1-\cos(n a)]\,[1-\cos(n-2)a]
\]
for $n\ge 2$, and employing the Cauchy--Schwarz inequality in the same
way as before, we find
\be
-1+{(r_1+r_{n-1})^2\over 1+r_{n-2}}\le r_n
\le 1-{(r_1-r_{n-1})^2\over 1-r_{n-2}} \;,
\elabel{rngeneralCS}
\ee
where the special case $n=3$ has been derived already. We have
thus found a set of inequalities for all correlation coefficients
$r_n$. In the next section, we will obtain bounds on the correlation
function using a different method, and will show that these ones are
stricter than those in (\ref{eq:rngeneralCS}).

\subsection{Constraints from a covariance matrix approach}
We will proceed in a
different way which is more straightforward. Consider a set of $N$ points
$x_m= m x$, with $m$ integer and $0\le m\le N-1$. The covariance matrix of the
random field at these $N$ points has the simple form
\be
\tens C_{ij}=\ave{g(i x) g(j x)}=\xi(|i-j|x) \;.
\ee
As is well known, the covariance matrix must be positive
semi-definite, i.e., its eigenvalues must be non-negative.
Dividing $\tens C$
by $\xi(0)>0$, we define
\be
\tens A_{ij}=\tens C_{ij}/\xi(0)=r_{|i-j|} \;,
\elabel{Adef}
\ee
and the
eigenvalues of $\tens A$ must obey $\lambda_i\ge 0$. For $N=2$, the
eigenvalues read $\lambda_{1,2}=1\pm r_1$, yielding
\be
\abs{r_1}\le 1\;,
\ee
i.e. we reobtain (\ref{eq:constr1}). For $N=3$, the eigenvalues read
\[
\lambda_1=1-r_2\;, \;\;\lambda_{2,3}={2+r_2\pm\sqrt{8
    r_1^2+r_2^2}\over 2} \;,
\]
and the conditions $\lambda_j\ge 0$ can be solved for $r_2$,
yielding (\ref{eq:constr2r}). The four eigenvalues of $\tens A$ in the case
$N=4$ are
\[
\lambda_{1,2,3,4}=1 \pm {1\over 2}\eck{r_1+r_3
\pm\sqrt{5 r_1^2-8 r_1 r_2+4 r_2^2-2 r_1 r_3+r_3^2}}\;,
\]
and the conditions $\lambda_i\ge 0$ after some algebra can be brought
into the form (\ref{eq:constr3r}). For $N\ge 5$, the eigenvalues of
$\tens A$ have a more complicated form; they are obtained as solutions of
higher-order polynomials (see below).

However, we do not need an explicit expression for the eigenvalues,
but only need to assure that they are non-negative. This condition can
be formulated in a different way. The eigenvalues of the matrix $\tens A$
are given by the roots of the characteristic polynomial, which is the
determinant of the matrix $\lambda \delta_{ij}-\tens A_{ij}$. For a given
$N$, this polynomial is of order $N$ in $\lambda$ and of the form
\be
\lambda^N+\sum_{k=0}^{N-1} b_k\lambda^k \;.
\elabel{charapoly}
\ee
The coefficients $b_k$ of the polynomial are functions of the $r_k$,
as obtained from calculating the determinant.
On the other hand, they can be expressed by the roots $\lambda_k$ of
the polynomial; for example, for $N=3$, one finds that
\[
\lambda^3+\sum_{k=0}^2 b_k\lambda^k
=\lambda^3-(\lambda_1+\lambda_2+\lambda_3)\lambda^2
+(\lambda_1\lambda_2+\lambda_1\lambda_3+\lambda_2\lambda_3)\lambda
-\lambda_1 \lambda_2 \lambda_3 \;.
\]
The condition that all $\lambda_k\ge 0$ is then equivalent to the
condition that $b_2\le 0$, $b_1\ge 0$, and $b_0\le 0$. It is easy to
show that these conditions lead to the constraint (\ref{eq:constr2r}),
together with $|r_1|\le 1$.

This procedure can be generalized for arbitrary $N$: the condition
that all eigenvalues of the correlation matrix $\tens A$ are non-negative is
equivalent to requiring that the coefficients of the characteristic
polynomial (\ref{eq:charapoly}) satisfy $b_{N-n}\le 0$ if $n$
is odd, and $b_{N-n}\ge 0$ if $n$ is even. The explicit calculation of
the characteristic polynomial by hand becomes infeasible, hence we use
the computer algebra program Mathematica (Wolfram 1996). For
successively larger $N$, we calculate the characteristic
polynomial. As we will show below, the
characteristic polynomial factorizes into two polynomials; if $N$ is even, it
factorizes into two polynomials of order $N/2$, whereas if $N$ is
odd, it factorizes into polynomials of order $(N\pm 1)/2$. The
condition that all eigenvalues are non-negative is equivalent to the
condition that the roots of both polynomials are non-negative; this
condition can be translated into inequalities for the coefficients of
the polynomials in the same way as described above.

We illustrate this procedure for the case $N=5$. The characteristic
polynomial in this case becomes
\be
\rund{\lambda^2+b_1\lambda+b_0}
\rund{\lambda^3+c_2\lambda^2+c_1\lambda+c_0}\;,
\ee
with
\bea
b_1&=&r_2+r_4-2\nonumber \\
b_0&=&(1-r_2)(1-r_4)-(r_1-r_3)^2 \nonumber \\
c_2&=&-3-r_2-r_4\nonumber \\
c_1&=&3(1-r_1^2)-2r_1 r_3-r_3^2+2r_4+r_2(2+r_4)-2 r_2^2\nonumber \\
c_0&=&(2 r_1^2-1-r_2)r_4+r_3^2+2 r_1 r_3(1-2r_2^2)+2r_2^2(1+r_2)
-r_2+r_1^2(3-4r_2)-1\nonumber
\eea
The conditions $b_1\le 0$, $c_2\le 0$ and $c_1\ge 0$ are satisfied,
irrespective of the value of $r_4$, owing to $|r_n|\le 1$ for all $n$,
and the inequalities obtained before for $r_n$, $n\le 3$. Thus, these
three inequalities do not yield additional constraints of $r_4$. Those
come from requiring $b_0\ge 0$ and $c_0\le 0$. Both coefficients are
linear in $r_4$, which allows us to write the conditions explicitly,
\be
-1+{r_1^2(1-4 r_2)+2 r_2^2(1+r_2)+2 r_1 r_3(1-2 r_2)+r_3^2
\over 1-2 r_1^2+r_2}
\le r_4 \le 1-{(r_1-r_3)^2\over (1-r_2)}
\elabel{constr4r}
\ee
Alternatively, one can use the function {\tt InequalitySolve} of
Mathematica for the five inequalities for the four coefficients, to
obtain the same result. 

We see that the upper bound in (\ref{eq:constr4r}) agrees with that
obtained from the Cauchy--Schwarz inequality -- see
(\ref{eq:rngeneralCS}) -- but the lower bounds differ. Hence, for
$n=4$ the bounds on $r_n$ derived from the two methods are different,
and that will be the case for all $n\ge 4$. This is not surprising,
since the Cauchy--Schwarz inequality does not make any statement on
how `sharp' the bounds are, i.e., whether the bounds can actually be
approached. One can argue, using discrete Fourier transforms, that the
bounds obtained from the covariance method are `sharp' (for a
one-dimensional field) in the sense that for each combination of
allowed $r_n$, one can find a non-negative power spectrum
corresponding to these correlation coefficients -- the bounds can
therefore not be narrowed further. 

It should be noted that in the cases given above, the lower (upper)
bound on $r_n$ is a quadratic function in $r_{n-1}$, and its quadratic
term $c r_{n-1}^2$ has a positive (negative) coefficient $c$. This
implies that the allowed region in the $r_{n-1}$-$r_n$-plane is
convex; indeed, as we will show below, the allowed $n$-dimensional
region for the $r_1, \dots, r_n$ is convex.

The procedure illustrated for the case $N=5$ holds for larger $N$ as
well: only the coefficients of $\lambda^0$ in the two polynomials
yield new constraints on the correlation coefficient $r_{N-1}$, and
these two coefficients are linear in $r_{N-1}$, so the constraints for
$r_{N-1}$ can be obtained explicitly. This then leads us to conclude
that the positive-definiteness of the matrix $\tens{A}$ for a given
$N$ is equivalent to the positivity of the determinants of all
submatrices which are obtained by considering only the first $n$ rows
and columns of $\tens{A}$, with $1\le n\le N$. We will make use of
this property in the next subsection.

For larger $N$, the upper and lower bounds are given by
quite complicated expressions, so we refrain from writing them down;
however, using the output from Mathematica, they can
be used for subsequent numerical calculations. A much
more convenient method for calculating the upper and lower bounds
explicitly will be obtained below.

To summarize, the procedure outlined gives us constraints on the form
\be
r_{n{\rm l}}\le r_n\le r_{n{\rm u}} \;,
\ee
where the lower and upper bounds are function of the $r_k$, $k<n$, and
satisfy $-1\le r_{n{\rm l}}\le r_{n{\rm u}}\le 1$.
For $n\le 4$, the bounds $r_{n{\rm l}}$ and $r_{n{\rm u}}$ have been
written down explicitly above.

The existence of these upper and lower bounds on $r_n$, and thus on
the correlation function, immediately implies that the likelihood
function of the correlation function cannot be a Gaussian, since the
support of the latter is unbounded. This does not imply necessarily
that the Gaussian approximation is bad; if the Gaussian probability
ellipsoid described by (\ref{eq:LLikeli}) is `small' in the sense that
it is well contained inside the allowed region, the existence of the
bounds alone yields no estimate for the accuracy of the Gaussian
approximation. We will come back to this issue in
Sect.\ts\ref{sc:sim}.

Whereas the expressions for $r_{n{\rm l}}$ and $r_{n{\rm u}}$ for
larger $n$ become quite long, we found a remarkable result for the
difference $\Delta_n\equiv r_{n{\rm u}}-r_{n{\rm l}}$. This is most
easily expressed by defining the functions
\be
{\cal F}_n(r_1,\dots,r_n)=(r_{n{\rm u}}-r_n)(r_n-r_{n{\rm l}})\;.
\ee
Then, for odd $n$
\be
\Delta_n(r_1,\dots,r_{n-1})=2 \rund{\prod_{k=1}^{(n-1)/2} {\cal F}_{2k}}
\rund{\prod_{k=1}^{(n-1)/2} {\cal F}_{2k-1}}^{-1}\;,
\elabel{Deltaodd}
\ee
and for even $n$,
\be
\Delta_n(r_1,\dots,r_{n-1})=2 \rund{\prod_{k=1}^{n/2}{\cal F}_{2k-1}}
\rund{\prod_{k=1}^{n/2-1}{\cal F}_{2k}}^{-1}\;.
\elabel{Deltaeven}
\ee
This result has been obtained by guessing, and subsequently verified
with Mathematica, up to $n=16$, using the explicit expressions for the
bounds derived next.  We will make use of these properties in the
Sect.\ts\ref{sc:3}.

\subsection{Explicit expression for the bounds}
We will now derive an explicit expression for the upper and lower
bounds on the $r_n$. In doing so, we will first show that the
determinant of the matrix $\tens A$ for $N$ points factorizes into two
polynomials, both of which are linear in $r_{N-1}$. Then we make use
of the fact that the positive definiteness of $\tens{A}$ is equivalent
to having positive determinant of all submatrices obtained from
$\tens{A}$ by considering only the first $n$ rows and columns, $n\le
N$; however, these submatrices of $\tens{A}$ correspond to the matrix
$\tens{A}$ for lower numbers of points, and their positive determinant
is equivalent to the upper and lower bounds on the $r_n$ for $n\le
N-2$. Hence, by increasing $N$ successively, the constraints on
$r_{N-1}$ are obtained from requiring $\det(\tens{A})\ge 0$. 

The determinant of a matrix is unchanged if a multiple of one row
(column) is added to another row (column). We make use of this fact
for the calculation of $\det \tens A$, by carrying out the following four
steps:
\ben
\item
We add the first row to the last, obtaining the matrix $\tens A^{(1)}$ with
elements
\[
\tens A^{(1)}_{ij}=\tens A_{ij}+\delta_{iN} \tens A_{1j}\;.
\]
\item
We subtract the last column from the first,
\[
\tens A^{(2)}_{ij}=\tens A^{(1)}_{ij}-\delta_{1j} \tens A^{(1)}_{iN}\;.
\]
\item
The second row is added to the $(N-1)$-st one, the third row is added
to the $(N-2)$-nd one, and so on. For $N$ odd, this reads
\[
\tens A^{(3)}_{ij}=\tens A^{(2)}_{ij} + \sum_{k=1}^{(N-3)/2} \delta_{(N-k)i}
\tens A^{(2)}_{(1+k)j} \;,
\]
whereas for $N$ even, the sum extends to $(N-2)/2$.

\item
Finally, the $(N-1)$-st column is subtracted from the second one, the
$(N-2)$-nd column is subtracted from the third one, and so on, which
for odd $N$ reads
\[
\tens A^{(4)}_{ij}=\tens A^{(3)}_{ij} - \sum_{k=1}^{(N-3)/2} \delta_{(1+k)j}
\tens A^{(3)}_{i(N-k)} \;,
\]
whereas for even $N$ the sum extends to $(N-2)/2$.
\een

\begin{figure*}[t]
\[
\tens A=\left(
                  \begin{array}{ccccccc}
                   1 & r_1 & r_2 & r_3 & r_4 & r_5 & r_6 \\
                   r_1 & 1 & r_1 & r_2 & r_3 & r_4 & r_5 \\
                   r_2 & r_1 & 1 & r_1 & r_2 & r_3 & r_4 \\
                   r_3 & r_2 & r_1 & 1 & r_1 & r_2 & r_3 \\
                   r_4 & r_3 & r_2 & r_1 & 1 & r_1 & r_2 \\
                   r_5 & r_4 & r_3 & r_2 & r_1 & 1 & r_1 \\
                   r_6 & r_5 & r_4 & r_3 & r_2 & r_1 & 1
                  \end{array}
                  \right)
\] \[
\tens A^{(1)}=  \left(
   \begin{array}{ccccccc}
    1 & r_1 & r_2 & r_3 & r_4 & r_5 & r_6 \\
    r_1 & 1 & r_1 & r_2 & r_3 & r_4 & r_5 \\
    r_2 & r_1 & 1 & r_1 & r_2 & r_3 & r_4 \\
    r_3 & r_2 & r_1 & 1 & r_1 & r_2 & r_3 \\
    r_4 & r_3 & r_2 & r_1 & 1 & r_1 & r_2 \\
    r_5 & r_4 & r_3 & r_2 & r_1 & 1 & r_1 \\
    r_6+1 & r_1+r_5 & r_2+r_4 & 2 r_3 & r_2+r_4 & r_1+r_5 & r_6+1
   \end{array}
   \right)
\]

\[
\tens A^{(2)}= \left(
   \begin{array}{ccccccc}
    1-r_6 & r_1 & r_2 & r_3 & r_4 & r_5 & r_6 \\
    r_1-r_5 & 1 & r_1 & r_2 & r_3 & r_4 & r_5 \\
    r_2-r_4 & r_1 & 1 & r_1 & r_2 & r_3 & r_4 \\
    0 & r_2 & r_1 & 1 & r_1 & r_2 & r_3 \\
    r_4-r_2 & r_3 & r_2 & r_1 & 1 & r_1 & r_2 \\
    r_5-r_1 & r_4 & r_3 & r_2 & r_1 & 1 & r_1 \\
    0 & r_1+r_5 & r_2+r_4 & 2 r_3 & r_2+r_4 & r_1+r_5 & r_6+1
   \end{array}
   \right)
\]
\[
\tens A^{(3)}= \left(
   \begin{array}{ccccccc}
    1-r_6 & r_1 & r_2 & r_3 & r_4 & r_5 & r_6 \\
    r_1-r_5 & 1 & r_1 & r_2 & r_3 & r_4 & r_5 \\
    r_2-r_4 & r_1 & 1 & r_1 & r_2 & r_3 & r_4 \\
    0 & r_2 & r_1 & 1 & r_1 & r_2 & r_3 \\
    0 & r_1+r_3 & r_2+1 & 2 r_1 & r_2+1 & r_1+r_3 & r_2+r_4 \\
    0 & r_4+1 & r_1+r_3 & 2 r_2 & r_1+r_3 & r_4+1 & r_1+r_5 \\
    0 & r_1+r_5 & r_2+r_4 & 2 r_3 & r_2+r_4 & r_1+r_5 & r_6+1
   \end{array}
   \right)
\]
\[
\tens A^{(4)}= \left(
   \begin{array}{ccccccc}
    1-r_6 & r_1-r_5 & r_2-r_4 & r_3 & r_4 & r_5 & r_6 \\
    r_1-r_5 & 1-r_4 & r_1-r_3 & r_2 & r_3 & r_4 & r_5 \\
    r_2-r_4 & r_1-r_3 & 1-r_2 & r_1 & r_2 & r_3 & r_4 \\
    0 & 0 & 0 & 1 & r_1 & r_2 & r_3 \\
    0 & 0 & 0 & 2 r_1 & r_2+1 & r_1+r_3 & r_2+r_4 \\
    0 & 0 & 0 & 2 r_2 & r_1+r_3 & r_4+1 & r_1+r_5 \\
    0 & 0 & 0 & 2 r_3 & r_2+r_4 & r_1+r_5 & r_6+1
   \end{array}
   \right)
\]

\caption{For $N=7$, the original covariance matrix $\tens A$ is shown,
  together with four transformations of it, described in the text,
  which leave the determinant unchanged}
\label{fig:matrix}
\end{figure*}

These steps are illustrated for the case $N=7$ in
Fig.\ts\ref{fig:matrix}.  For $N$ even [odd], this results in a matrix
$\tens A^{(4)}$ which contains in the lower left corner a $N/2\times
N/2$ [$(N-1)/2\times (N+1)/2$] submatrix with elements
zero. Therefore, the determinant of $\tens A^{(4)}$ factorizes into
the determinants of two $N/2\times N/2$ matrices [of a $(N-1)/2\times
(N-1)/2$ and of a $(N+1)/2\times (N+1)/2$ matrix].  Thus,
\be
\det(\tens A)\equiv \det\rund{\tens A^N} = \det\rund{\tens A^{(4)}}
=\det\rund{\tens U^N}\,\det\rund{\tens V^N}\;,
\ee
where we explicitly indicate the number $N$ of points, and thus the
dimensionality of $\tens A$, with a superscript,
and
where for even $N$, $\tens U^N$ and $\tens V^N$ are $N/2\times N/2$ matrices with
elements
\be
\tens U^N_{ij}=r_{|i-j|}-r_{N+1-i-j} \; ;\quad
\tens V^N_{ij}=r_{|i-j|}+r_{i+j-1}\;,\quad 1\le i,j\le N/2\;,
\ee
whereas for odd $N$, $\tens U^N$ and $\tens V^N$ are $(N-1)/2\times (N-1)/2$ and
$(N+1)/2\times (N+1)/2$ matrices, respectively, with elements
\be
\tens U^N_{ij}=r_{|i-j|}-r_{N+1-i-j}\;, \quad 1\le i,j\le (N-1)/2\;;
\quad
\tens V^N_{ij}=r_{|i-j|}+\rund{1-\delta_{1i}}r_{i+j-2}\;,
\quad 1\le i,j\le (N+1)/2\; .
\ee
Since $\tens U^N_{11}=1-r_{N-1}$, and the $(N/2-1)\times (N/2-1)$ (for $N$ even;
for $N$ odd this is a [$(N-1)/2-1]\times [(N-1)/2-1$]) submatrix obtained
by cancelling the first column and row of $\tens U^N$ is just $\tens U^{N-2}$, we
can write
\be
\det\rund{\tens U^N}=(1-r_{N-1})\det\rund{\tens U^{N-2}}+\det\rund{\bar{\tens{U}}^N}\;,
\elabel{detUN}
\ee
where $\bar{ \tens U}^N$ is the matrix which is obtained from $\tens
U^N$ by setting $\bar{ \tens U}^N_{11}=0$. The upper bound for
$r_{N-1}$ is found by setting $\det\rund{\tens U^N}\ge0$, which yields
\be
r_{n}\le 
r_{n{\rm u}}=1+{\det\rund{\bar{ \tens U}^{n+1}}
\over \det\rund{\tens U^{n-1}}}\;,
\ee
where we set $n=N-1$.
Analogously, the final element of $\tens V^N$ reads
$\tens V^N_{mm}=1+r_{N-1}$, where $m=N/2$ for
even $N$, and $m=(N+1)/2$ for odd $N$. Therefore,
\be
\det\rund{\tens V^N}=(1+r_{N-1})\det\rund{\tens V^{N-2}}+\det\rund{\bar{ \tens V}^N}\;,
\ee
where $\bar{ \tens V}^N$ is obtained from $\tens V^N$ by setting $\tens V_{mm}=0$; the
lower bound for $r_{N-1}$ is then obtained by setting this expression
to zero, or
\be
r_{n{\rm l}}=-1-{\det\rund{\bar{\tens V}^{n+1}}\over \det\rund{\tens V^{n-1}}}\;.
\ee
Since $\det\rund{\tens U^N}$ is a linear function of $r_{N-1}$, it must be
of the form $\det\rund{\tens U^N}=c(r_{N-1}-d)$, where the coefficients $c,
d$ are independent of $r_{N-1}$. The value of $d$ yields the root of
$\det\rund{\tens U^N}$, and is thus the upper bound on $r_{N-1}$. The
coefficient $c$ follows from (\ref{eq:detUN}); therefore,
\be
\det\rund{\tens U^{n+1}}=\rund{r_{n{\rm u}}-r_n} \det\rund{\tens U^{n-1}}\;.
\ee
The analogous result holds for the $\tens V^N$, i.e.
\be
\det\rund{\tens V^{n+1}}=\rund{r_n-r_{n{\rm l}}} \det\rund{\tens V^{n-1}}\;.
\ee
These recursion relations then yield the explicit expressions
\be
\det\rund{\tens U^N}=\prod_{k=1}^{N/2} \rund{r_{(2k-1){\rm u}}-r_{2k-1}}\; ;\quad
\det\rund{\tens V^N}=\prod_{k=1}^{N/2} \rund{r_{2k-1}-r_{(2k-1){\rm l}}}
\ee
for $N$ even, and
\be
\det\rund{\tens U^N}=\prod_{k=1}^{(N-1)/2} \rund{r_{(2k){\rm u}}-r_{2k}}\; ;\quad
\det\rund{\tens V^N}=\prod_{k=1}^{(N-1)/2} \rund{r_{2k}-r_{(2k){\rm l}}}
\ee
for $N$ odd. This yields for the determinant of the matrix $\tens A^N$ the
explicit expression
\be
\det\rund{\tens A^N}=\sum_{k=1}^{N/2} {\cal F}_{2k-1}\;,\quad
\det\rund{\tens A^N}=\sum_{k=1}^{(N-1)/2} {\cal F}_{2k}
\ee
for even and odd $N$, respectively.
Accordingly, the width $\Delta_n=r_{n{\rm u}}-r_{n{\rm
    l}}$ of the allowed range of $r_n$ then becomes
\be
\Delta_n=2 {\det\rund{\tens A^N}\over \det\rund{\tens A^{N-1}}}\;,
\ee
where we made use of (\ref{eq:Deltaodd}) and (\ref{eq:Deltaeven}).

\section{\llabel{3}How strong are the constraints?}
We shall now investigate the question how constraining the constraints
on the correlation function or, equivalently, the correlation
coefficients are. In order to quantify this question, we imagine that
the correlation function is measured on a regular grid of separations
$n x$, and from that we define as before $r_n=\xi(n x)/\xi(0)$. Due to
(\ref{eq:constr1}), $|r_n|\le 1$ for all $n$. We therefore consider the set of
functions defined by their values on the grid points, and allow only
values in the interval $-1\le r_n\le 1$ for all $n$. Let $M$ be the
largest value of $n$ that we consider; then the functions we consider
are defined by a point in the $M$-dimensional space spanned by the
$r_n$. This $M$-dimensional cube has sidelength 2 and thus a volume of
$2^M$. We will now investigate which fraction of this volume
corresponds to correlation functions which satisfy the constraints
derived in the previous section.

A related question that can be answered is: suppose the values of
$r_k$, $1\le k\le n$ are given; what fraction of the
$(M-n)$-dimensional subspace, spanned by the $r_k$ with $n+1\le k\le
M$ corresponds to allowed correlation functions? For example, if
$r_1=1$, then all other $r_k=1$, and hence the condition $r_1=1$ is
very constraining -- the volume fraction of the subspace spanned by
the $r_k$ with $k\ge 2$ is zero in this case.

Mathematically, we define these volume fractions as
\bea
V_{Mn}&=&\int_{r_{(n+1){\rm l}}}^{r_{(n+1){\rm u}}}{\d r_{n+1}\over 2}
\int_{r_{(n+2){\rm l}}}^{r_{(n+2){\rm u}}}{\d r_{n+2}\over 2} \dots
\int_{r_{M{\rm l}}}^{r_{M{\rm u}}}{\d r_M\over 2} \nonumber \\
&=&\prod_{k=n+1}^M \int_{r_{k{\rm l}}}^{r_{k{\rm u}}}{\d r_k\over 2} \;.
\eea
The factor $1/2$ in each integration accounts for the side-length of
the $(M-n)$-dimensional cube, so that the $V_{Mn}$ are indeed fractional
volumes. $V_{Mn}$ depends on the $r_k$ with $k\le n$; in particular,
$V_{M0}\equiv V_M$ is the fractional volume which is allowed if all
constraints are taken into account. From the definition of the
$V_{Mn}$ it is obvious that the following recursion holds:
\be
V_{Mn}=\int_{r_{(n+1){\rm l}}}^{r_{(n+1){\rm u}}}{\d r_{n+1}\over 2}
V_{M(n+1)}\;.
\ee
We can therefore calculate the $V_{Mn}$ iteratively, starting with
\be
V_{M(M-1)}=\int{\d r_M\over 2}={\Delta_M\over 2}
={{\cal F}_{M-1}{\cal F}_{M-3}\dots \over
{\cal F}_{M-2}{\cal F}_{M-4}\dots}\;,
\ee
where we have skipped the integration limits; here
and in the following, an integral over $r_n$ always extends from
$r_{n{\rm l}}$ to $r_{n{\rm u}}$. Here we made use of
(\ref{eq:Deltaodd}) or (\ref{eq:Deltaeven}), depending on $M$. The
dots indicate that factors are added until ${\cal F}_1$ or ${\cal
F}_2$ is reached.

\begin{figure}[h!]
%\hspace{-1cm}
\resizebox{0.5\hsize}{!}{\includegraphics{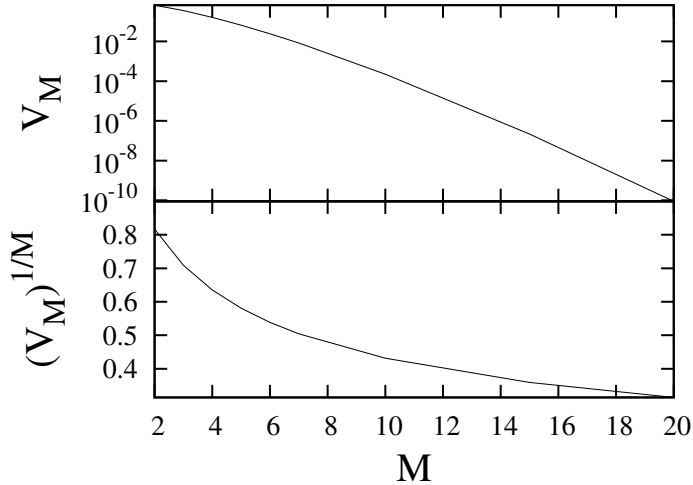}}
\caption{Upper panel: volume fraction $V_M$ (Eq.~\ref{eq:VMdef}) as
  function of the number $M$ of separations for which the correlation
  function is measured. Lower panel: $V_M^{1/M}$ as function of
  $M$. One sees that the typical linear dimension of the allowed
  region decreases with $M$, meaning that the strong decrease of $V_M$
with $M$ is not just an effect of the dimensionality of the volume
considered }
\label{fig:volfrac}
\end{figure}

It should be noted that the only dependence of $V_{M(M-1)}$ on
$r_{M-1}$ is through the factor ${\cal F}_{M-1}$. Therefore, the next
recursion step reads
\bea
V_{M(M-2)}&=&\int{\d r_{M-1}\over 2}\;V_{M(M-1)} \nonumber \\
&=&{1\over 2} {{\cal F}_{M-3}{\cal F}_{M-5}\dots \over
{\cal F}_{M-2}{\cal F}_{M-4}\dots}
\int \d r_{M-1}\; {\cal F}_{M-1}  \nonumber \\
&=&{1\over 2} {{\cal F}_{M-3}{\cal F}_{M-5}\dots \over
{\cal F}_{M-2}{\cal F}_{M-4}\dots}\,\Delta_{M-1}^3\,{\rm B}(2,2)
\nonumber \\
&=&2^2 {\rm B}(2,2) \rund{{\cal F}_{M-2}{\cal F}_{M-4}\dots \over
{\cal F}_{M-3}{\cal F}_{M-5}\dots }^2 \;,
\eea
where ${\rm B}(x,y)=\Gamma(x)\,\Gamma(y)/\Gamma(x+y)$ is the
beta-function, and we used the relation
\be
\int_a^b \d x\;(b-x)^n (x-a)^n = (b-a)^{1+2n}\,{\rm B}(1+n,1+n)\;.
\elabel{betaintegral}
\ee
For the next step we notice that the only dependence of $V_{M(M-2)}$
on $r_{M-2}$ is through the function ${\cal F}_{M-2}$, which therefore
lets us write
\bea
V_{M(M-3)}&=&\int{\d r_{M-2}\over 2}\;V_{M(M-2)} \nonumber \\
&=&
2 {\rm B}(2,2)\rund{{\cal F}_{M-4}{\cal F}_{M-6}\dots \over
{\cal F}_{M-3}{\cal F}_{M-5}\dots }^2
\int{\d r_{M-2}\over 2}\;{\cal F}_{M-2}^2\nonumber \\
&=&
2^6\, {\rm B}(2,2)\,{\rm B}(3,3)\rund{{\cal F}_{M-3}{\cal F}_{M-5}\dots
\over {\cal F}_{M-4}{\cal F}_{M-6}\dots}^3 \;,
\eea
where we made again use of (\ref{eq:betaintegral}).
Based on these results, we can now obtain a general expression,
\be
V_{M(M-n)}=2^{n(n-1)}\eck{\prod_{k=2}^n{\rm B}(k,k)}
\rund{{\cal F}_{M-n}{\cal F}_{M-n-2}\dots \over
{\cal F}_{M-n-1}{\cal F}_{M-n-3}\dots }^n\;,
\ee
which can be proved by induction. In particular, the $V_M\equiv
V_{M0}$ are given as
\be
V_M=2^{M(M-1)}\eck{\prod_{k=2}^M {\rm B}(k,k)} \;.\elabel{VMdef}
\ee
The values of $V_M$ up to $M=20$ are shown in
Fig.~\ref{fig:volfrac}. As can be seen from the upper panel, the
admissible volume very quickly decreases as $M$
increases. For comparison, in the lower panel we show $V_M^{1/M}$,
i.e.~the typical diameter of the allowed region.

%
% The first few values of the $V_M$ are
% \bea
% v_2&=&{2\over 3}\approx 0.667 \nonumber \\
% v_3&=&{16\over 45} \approx 0.356 \nonumber \\
% v_4&=&{256 \over 1575} \approx 0.163 \nonumber \\
% v_5&=&{32768\over 496125} \approx 6.60 \times 10^{-2} \nonumber \\
% v_6&=&{8388608\over 343814625}\approx 2.44 \times 10^{-2} \nonumber \\
% v_7&=&{8589934592\over 1032475318875} \approx 8.32 \times 10^{-3}
% \nonumber \\
% v_{10}&\approx& 2.25 \times 10^{-4}\nonumber \\
% v_{15}&\approx&  2.15 \times 10^{-7}\nonumber \\
% v_{20}&\approx&  8.93 \times 10^{-11}\nonumber
% \eea

\section{\llabel{4}Generalizations}
The considerations of the previous sections were restricted to
correlation functions as measured at equidistant points $x_n=n x$. In
some cases, however, it may be advantageous to drop this constraint,
e.g., to consider the correlation function at logarithmically spaced
points. Furthermore, as mentioned before, tighter constraints on the
correlation function are expected to hold in the multi-dimensional
case. We shall consider these aspects in this section, presenting
another method for deriving constraints which yields the optimal
constraints for arbitrarily spaced points $x_n$ in one- or
higher-dimensional fields. But before, we briefly consider the covariance
method for higher dimensions.

\subsection{Higher-dimensional fields: the covariance method}
As was noted above, the inequalities for the correlation functions
have been obtained with a one-dimensional random field in
mind. Whereas we have shown that all the bounds on correlation
functions are also valid for higher-dimensional fields, they are not
assumed to be `optimal' -- the reason is that the equivalent
one-dimensional power spectrum defined in (\ref{eq:xiofP0}) was
assumed to be totally arbitrary, except from being non-negative,
whereas for isotropic fields in higher dimensions, it will obey
further constraint relations due to the $(n-1)$-dimensional
integration in (\ref{eq:mdto1d}). A first indication that the bounds
can be improved has been seen with the lower bounds on the ratio
$\xi(x)/\xi(0)$, which turned out to be larger for two and three
dimensions than for a one-dimensional field.  The foregoing
constraints on the correlation function are re-obtained with the
covariance matrix approach in higher dimensions if the set of points
are placed equidistant along one direction. New (and stronger)
constraints are expected from this method if the distribution of
points makes use of these higher dimensions.

As a first example, we consider a two-dimensional (or
higher-dimensional) field and place three points in an equilateral
triangle of side-length $x$. The separation between any pair of points
is then $x$, and the covariance matrix of these three points,
normalized by $\xi(0)$, then reads
\[
\tens A=\left(
   \begin{array}{ccc}
   1 & r_1 & r_1 \\
   r_1 & 1 & r_1 \\
   r_1 & r_1 & 1   \end{array}
   \right)  \;.
\]
The eigenvalues of this matrix are $\lambda_{1,2}=1-r_1$,
$\lambda_3=1+2 r_1$, and requiring their non-negativity leads to
\be
-1/2\le r_1=\xi(x)/\xi(0)\le 1
\elabel{rconstr2D1}
\ee
for all $x$. The lower bound is somewhat smaller than the one obtained
earlier for two-dimensional random fields, $\xi(x)/\xi(0)\gtrsim
-0.403$, but significantly larger than that obtained for the 1-D case,
$\xi(x)/\xi(0)\ge -1$. Next we consider a set of three points forming
a triangle of which two sides have length $x$, and the third side has
length $\eta x$, with $0\le \eta\le 2$. The corresponding covariance
matrix reads
\[
\tens A=\left(
   \begin{array}{ccc}
   1 & r_1 & r_1 \\
   r_1 & 1 & r_\eta \\
   r_1 & r_\eta & 1   \end{array}
   \right)  \;,
\]
and its eigenvalues are $\lambda_1=1-r_\eta$,
$\lambda_{2,3}=\rund{2+r_\eta\pm\sqrt{8 r_1^2+r_\eta^2}}/2$; note that
we used the notation $r_\eta=\xi(\eta x)/\xi(0)$. Non-negativity of
the eigenvalues leads to the constraints
\be
\max\rund{-1/2,2r_1^2-1}\le r_\eta\le 1\; {\rm for}\; 0\le \eta\le 2\;,
\elabel{rconstr2Deta}
\ee
where we also used (\ref{eq:rconstr2D1}).

Finally, we consider a square of side-length $x$, for which the
covariance matrix reads
\[
\tens A=\left(
   \begin{array}{cccc}
   1 & r_1 & r_1 & r_{\sqrt{2}}\\
   r_1 & 1 & r_{\sqrt{2}} & r_1 \\
   r_1 & r_{\sqrt{2}} & 1 & r_1 \\
   r_{\sqrt{2}} & r_1 & r_1 & 1  \end{array}
   \right)  \;,
\]
with the eigenvalues $\lambda_{1,2}=1-r_{\sqrt{2}}$,
$\lambda_{3,4}=1+r_{\sqrt{2}}\pm 2 r_1$. Their non-negativity, combined
with  (\ref{eq:rconstr2D1}), yields
\be
\max\rund{-1/2,\abs{r_1}-1}\le r_{\sqrt{2}}\le 1\; ,
\ee
which is a weaker constraint than (\ref{eq:rconstr2Deta}) for
$\eta=\sqrt{2}$. Thus we see that the choice of the geometrical
configuration of points affects the resulting constraints on the
correlation function. It is by no means clear how to find a
set of configurations such as to obtain the `optimal' constraints in
two (or higher) dimensions. In fact, methods other than
using the covariance matrix need to be considered for obtaining
optimal constraints (see below). 

Finally, we consider the simplest case in three dimensions, namely a
set a of four points arranged in a regular tetrahedron of sidelength
$x$. The resulting covariance matrix is
\[
\tens A=\left(
   \begin{array}{cccc}
   1 & r_1 & r_1 & r_1\\
   r_1 & 1 & r_1 & r_1 \\
   r_1 & r_1 & 1 & r_1 \\
   r_1 & r_1 & r_1 & 1  \end{array}
   \right)  \;,
\]
with eigenvalues $\lambda_{1,2,3}=1-r_1$, $\lambda_4=1+3r_1$, yielding
\be
-1/3\le r_1\le 1\;,
\ee
a constraint stronger than the ones obtained for one and two
dimensions, but falling short of the one derived from the global
minimum of the spherical Bessel function, $r_1\gtrsim -0.217$.

\subsection{Optimal constraints in the general case}
We now describe a general method for deriving constraints on
correlation functions $\xi(x_n)$ of homogeneous and isotropic random
fields, allowing for arbitrary values of separations $x_n$ and
arbitrary dimensions. However, it should be said right at the
beginning that we were unable to obtain the corresponding bounds on
the correlation functions explicitly.

We can write the general relation (\ref{eq:xiofP}) in the form
\be
\xi(x)=\int_0^\infty \d k\; \hat P(k)\, u(x k)\;,
\ee
where $\hat P(k)$ is non-negative and $u(0)=1$. For a 1-dimensional field,
$\hat P(k)=P(k)/(2\pi)$, $u(y)=\cos y$; for two dimensions,
$\hat P(k)=k\,P(k)/(2 \pi)$, $u(y)={\rm J}_0(y)$, and for three
dimensions, $\hat P(k)=k^2\,P(k)/(2 \pi^2)$, $u(y)={\rm j}_0(y)$. Next
we consider a quadrature formula for the integral, and write
\be
\xi(x)=\sum_{j=1}^K w_j\,\hat P(k_j)\,u(x k_j)
\equiv \sum_{j=1}^K W_j\,u(x k_j)\;,
\ee
where the $w_j$ are (positive) weights corresponding to the quadrature
formula, and in the last step we defined $W_j\ge 0$. This
approximation can be made arbitrarily accurate by letting
$K\to\infty$. Defining the correlation coefficient as before, we
obtain
\be
r(x)\equiv \xi(x)/\xi(0)=\sum_{j=1}^K V_j\,u(x k_j)\;,
\ee
where the coefficients
\be
V_i=\eck{\sum_{j=1}^K W_j}^{-1} W_i
\ee
satisfy $0\le V_i\le 1$ and $\sum V_i=1$.

If we now consider a set of $N$ points $x_n$, together with the
correlation coefficients $r_n\equiv r(x_n)$, then we see that a point
in the $N$-dimensional space of $\vec r=(r_1,\dots,r_N)$ can be
described as a weighted sum of points lying along the curve
$\vec c(\lambda)=(u(\lambda x_1), \dots,u(\lambda x_N))$, $0\le \lambda <
\infty$, i.e.,
\be
\vec r=\sum_{j=1}^K V_j\,\vec c(\lambda_j) \;,
\ee
where we considered the transition of the discrete points $k_j$ to a
continuous variable $\lambda$. 
Since $V_i\in[0,1]$ and $\sum V_i=1$, the point $\vec r$ must be
located inside the convex volume containing all points on the curve
$\vec c(\lambda)$.  It is clear that this convex envelope of the curve
$\vec c$ yields indeed the optimal general bounds on the $r_i$: every
point within the convex envelope can be realized by choosing a set of
$n$ points on the curve $\vec c$ appropriately. Since the function
$u(y)$ depends on the dimension of the random field, the constraints
will be different for different numbers of dimensions. Furthermore,
the curve $\vec c(\lambda)$ depends on the choice of the $x_n$, and
therefore the constraints will also depend on the choice of
separations for which the correlation function is measured.

Unfortunately, we have not found a way how to algebraically describe
the convex envelope of the curve $\vec c(\lambda)$, and thus to obtain
explicit expressions for the upper and lower bounds on $r_i$. For now,
we therefore will present just a few simple examples.

For the 1-dimensional case with two points $x_2= 2 x_1$, the curve
reads $\vec c(\lambda)=(\cos \lambda,\cos 2\lambda)$, with
$\lambda=x_1 k$; the same set of points is described by the
curve $\vec c'(a)=(a,2 a^2-1)$ (using trigonometric identities), $-1\le a\le 1$. Thus, the convex envelope
of $\vec c'$ is the region between the parabola $\vec c'$ and the line
$r_2=+1$, and thus we re-obtain the bounds (\ref{eq:constr2}). For the
choice $x_2=3 x_1$, the set of points of $\vec c(\lambda)$ is
equivalent to that of the curve $\vec c'(a)=(a,4 a^3-3a)$,
$a\in [-1,1]$. The convex envelope can then be described by the bounds
${\rm max}(-1,4 r_1^3-3 r_1)\le r_2\le {\rm min}(1,4 r_1^3-3 r_1)$, where the
lower bound differs from $-1$ for $r_1> 1/2$, and the upper bound is
different from $+1$ for $r_1< -1/2$. If we
choose instead $x_2=\mu x_1$, then for a generic (non-rational) $\mu$,
the curve $\vec c(\lambda)$ fills the whole square $-1\le r_1\le 1$,
$-1\le r_2\le 1$, which is also coincident with its convex envelope.

\begin{figure}
\hspace{-1.3cm}
\begin{minipage}{0.32\textwidth}
\resizebox{1.5\hsize}{!}{\includegraphics[angle=270]{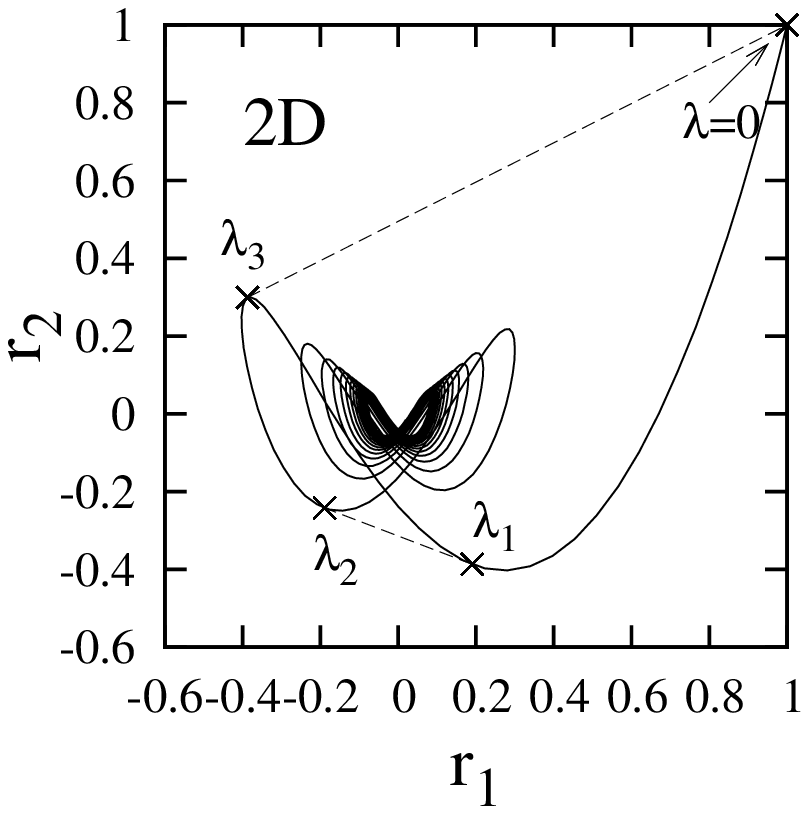}}
\end{minipage}
% \hspace{0.04\textwidth}
\begin{minipage}{0.32\textwidth}
\resizebox{1.5\hsize}{!}{\includegraphics[angle=270]{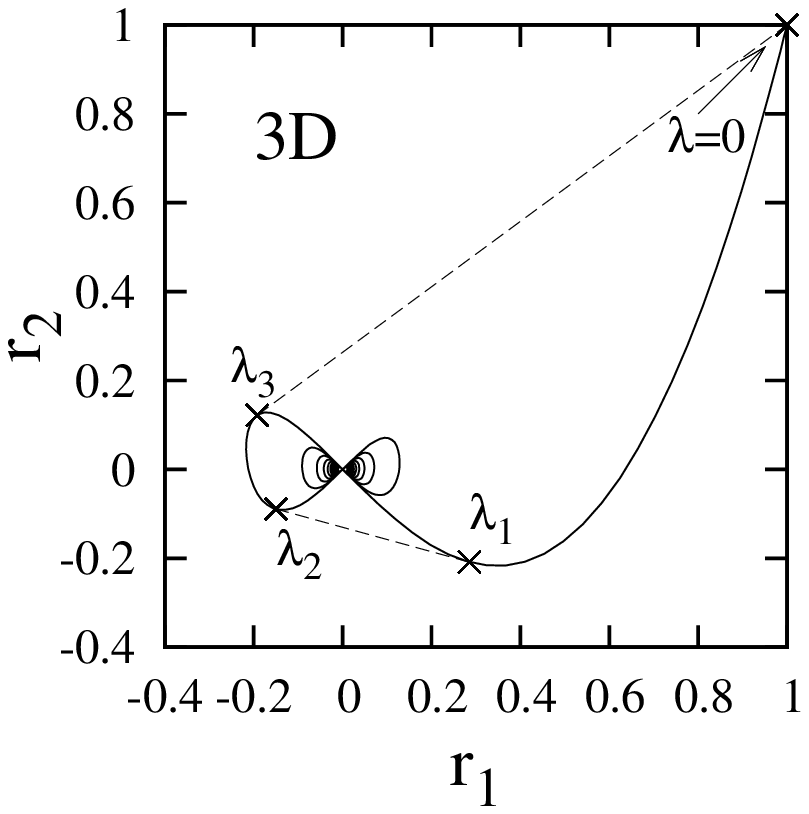}}
\end{minipage}
\begin{minipage}{0.32\textwidth}
\resizebox{1.5\hsize}{!}{\includegraphics[angle=270]{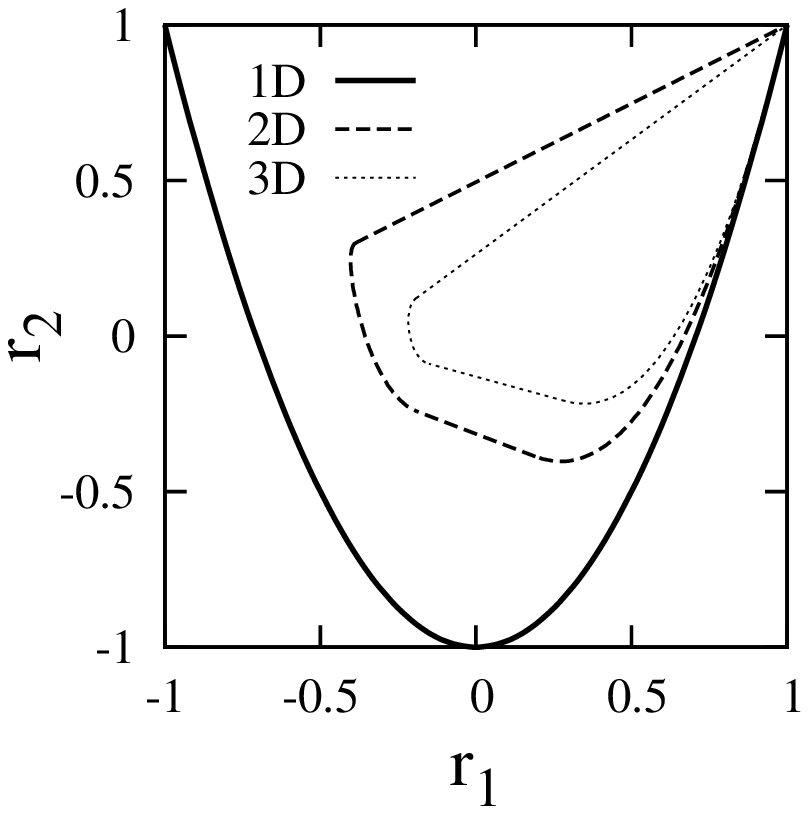}}
\end{minipage}
%\resizebox{0.95\hsize}{!}{\includegraphics[width=0.5\textwidth,angle=0]{bess1.eps}}
\caption{Left panel: The curve $\vec c(\lambda)=({\rm
  J}_0(\lambda),{\rm J}_0(2\lambda))$, together with its convex
envelope. The latter consists of two segments of the curve $\vec c$
and two straight lines, one of which is tangent to $\vec c(\lambda)$
at the points  $\vec c(\lambda_1)\approx(0.1905,-0.3870)$ and $\vec
c(\lambda_2)\approx(-0.1894,-0.2421)$, the other one is tangent to $\vec c$
at $\vec c(\lambda_3)\approx(-0.3872,0.2987)$ and intersects $\vec
c(0)=(1,1)$. The corresponding values of the curve parameter are
$\lambda_1\approx 2.0580$, $\lambda_2\approx 4.9636$, and
$\lambda_3\approx 3.5561$. These numerical values are found as
follows: the requirement that the lower straight line segment is
tangent to curve at the two points $\lambda_{1,2}$ leads to the
conditions $\dot{\vec c}(\lambda_i)=a_i\eck{\vec c(\lambda_1)-\vec
  c(\lambda_2)}$ for $i=1,2$, where $a_i$ are scalars. These four
  scalar equations for the four unknowns $a_i$, $\lambda_i$ have
  several solutions, the relevant one is the `outermost' one. For the
  upper straight line segment, one employs the condition that the
  tangent at point $\lambda_3$ goes through $\vec c(0)$, i.e.,
$\dot{\vec c}(\lambda_3)=a\eck{\vec c(\lambda_3)-\vec
  c(0)}$, and of the multiple solutions of these two scalar equations
for the two unknowns $a$ and $\lambda_3$, one takes the `outermost'
one. Middle panel: in a similar way, the curve $\vec c(\lambda)=({\rm
  j}_0(\lambda),{\rm j}_0(2\lambda))$ is plotted, together with its
convex envelope. Here, $\lambda_1\approx 2.3911$,
$\lambda_2\approx5.3490$ and
$\lambda_3\approx 4.0287$, and $\vec c(\lambda_1)\approx(0.2852,-0.2086)$,
$\vec c(\lambda_2)\approx(-0.1503,-0.0894)$, and
$\vec c(\lambda_3)\approx(-0.1924,0.1216)$. Right panel: Comparison of the allowed regions in the $r_1$-$r_2$-plane in 1, 2 and 3 dimensions.}
\label{fig:Bess}
\end{figure}

As the next example, we consider a 2-dimensional field with $x_2=2
x_1$. In the left panel of Fig.\ts\ref{fig:Bess}, the curve $\vec c(\lambda)=({\rm
  J}_0(\lambda),{\rm J}_0(2\lambda))$ is plotted for $\lambda \ge
0$. The convex envelope in this case can be constructed explicitly:
The boundary of the smallest convex region which contains the curve
$\vec c$ is composed of four parts: (1) The section of the curve $\vec
c(\lambda)$ for $0\le \lambda\le \lambda_1$, (2) the straight line
connecting the two points $\vec c(\lambda_1)$ and $\vec c(\lambda_2)$,
(3) the section of the curve $\vec c(\lambda)$ for $\lambda_3\le
\lambda\le\lambda_2$, and (4) the straight line connecting $\vec
c(\lambda_3)$ and $\vec c(0)=(1,1)$. In a similar way, the convex
envelope can be constructed for a 3-dimensional random field; see
Fig.\ts\ref{fig:Bess}, middle panel.

Unfortunately, we have not yet found a systematic way how to construct
the convex envelope for $n>2$, and how to obtain explicit bounds on
the correlation coefficients in these cases -- although it is clear
that the allowed region, at least in the $r_1$-$r_2$-plane, decreases
as one goes to higher-dimensional fields (see right panel of
Fig.\ts\ref{fig:Bess}). Therefore, the development of explicit bounds
in higher dimensions is of great interest.

\section{\llabel{5}Transformation of variables}
The finite bounds on the correlation coefficients clearly show that
the likelihood of the correlation function cannot be a
Gaussian. However, the bounds on $r_n$ may suggest that the Gaussian
approximation for the likelihood could be better in terms of
transformed variables, in which the allowed range for each $r_n$ is
mapped onto the real axis. Such a transformation would simplify the
parametrization of the likelihood from numerical simulations.
Defining
\be
x_n:={2 r_n-r_{n{\rm u}}-r_{n{\rm l}} \over
r_{n{\rm u}}-r_{n{\rm l}}}\;,
\elabel{rntoxn}
\ee
the allowed range of $r_n$ is mapped onto $-1<x_n<1$. It
should be noted that (\ref{eq:rntoxn}) is a coupled non-linear
transformation of variables, since the bounds on $r_n$ depend on the
$r_k$, $k<n$. The mapping to the real axis is then obtained by any
function that maps $[-1,1]$ to $(-\infty,\infty)$; we choose
\be
y_n:={\rm atanh} (x_n)\;.\elabel{xntoyn}
\ee
Thus, no bounds on the correlation coefficients are violated if the
probability distribution of the $y_n$ would follow a multi-variate
Gaussian.

We next consider the relation between the likelihood of the
correlation function and the corresponding likelihood of the $y_n$. To
shorten notation, we drop the explicit dependence on the model
parameters $p$ in the following. Then,
\bea
{\cal L}_\xi(\xi_0,\xi_1,\dots,\xi_N) \prod_{k=0}^N \d \xi_k
&=&{\cal L}'(\xi_1,\dots,\xi_N|\xi_0) \prod_{k=1}^N \d \xi_k \;
{\cal L}_0(\xi_0)\,\d \xi_0
={\cal L}_r(r_1,\dots,r_N|\xi_0) \prod_{k=1}^N \d r_k  \;
{\cal L}_0(\xi_0)\,\d \xi_0 \nonumber \\
&=&{\cal L}_y(y_1,\dots,y_N|\xi_0)\prod_{k=1}^N \d y_k  \;
{\cal L}_0(\xi_0)\,\d \xi_0\;,
\eea
where in the first step we have used the product rule of probability theory and introduced the
conditional probability density ${\cal L}'$, and where the next two
steps define the likelihood in terms of the $r_n$ and the $y_n$,
respectively. Thus, the distribution of the $r_n$ and $y_n$ can depend
explicitly on the value of $\xi_0$. 

The relation between ${\cal L}_r$ and ${\cal L}_y$ is obtained from
\be
{\cal L}_r(r_1,\dots,r_N|\xi_0)
={\cal L}_y(y_1,\dots,y_N|\xi_0) \det(J)\;,
\ee
where the $y_i$ are functions of the $r_j$, and
the transformation matrix $J$ is given by
\be
J_{ij}\equiv {\partial y_i\over \partial r_j}=
{1\over 1-x_i^2}\,{\partial x_i\over \partial r_j}
={2\over \Delta_i^2-(2 r_i-r_{i{\rm u}}-r_{i{\rm l}})^2}
\eck{\Delta_i \delta_{ij}-(r_i-r_{i\rm l}){\partial r_{i\rm
      u}\over \partial r_j}
-(r_{i\rm u}-r_i){\partial r_{i\rm
      l}\over \partial r_j}} \;.
\ee
Note that the partial derivatives vanish for $j\ge i$.
Thus, the transformation between the probability distribution of the
$r_i$ and that of the $y_i$ is rather complicated, implying that the
behaviour of these two probability distributions can be considerably
different. In particular, the ${\cal L}_y$ could be approximated by a
Gaussian, the corresponding ${\cal L}_r$ would have a significantly
different functional form; also the covariances of the two
distributions will differ significantly.

\section{\llabel{6}Simulations} \label{sc:sim}
In order to illustrate the analytical results discussed in the
previous sections and to explore the effects of the constraints on the
shape of the likelihood of the correlation function, we have conducted
some simple numerical experiments. We have generated realizations of a
periodic one-dimensional Gaussian random field $\delta$ on a regular
grid according to
\be
	g_i = \frac{1}{N}\sum_{j=0}^N\,{\rm e}^{2\pi{\rm i}\,
          ij/N}\,\tilde g_j\;,
\elabel{rfdef}
\ee
where $\tilde g_i$ is the discrete Fourier transform of the random
field, and has a normal distribution of zero mean, $\tilde g_i
\sim \mathcal{N}\left(0,\sqrt{N\,P_i}/2\right)$, where $P_i$ is the discretized
power spectrum and  $\mathcal{N}(\mu,\sigma)$ is the Gaussian distribution with mean $\mu$ and standard deviation $\sigma$.  From each realization, we measure the correlation
function using the estimator
\be
	\hat\xi_i =\frac{1}{N}  \sum_{a=0}^N\,g_a\,g_{a+i} \elabel{cfest}
\ee
and calculate the correlation coefficients $r_i$. Note that this
estimator of the correlation function explicitly employs the
periodicity of the discrete random field. As defined in this way, the
estimated correlation function is the Fourier transform of a power
spectrum $P'_i$, proportional to the squares of the amplitudes of the
$\tilde g_i$. Hence, $P'_i$ is non-negative, and we thus expect
that the correlation function of each realization obeys the
constraints derived before -- indeed, this is the case.

\begin{figure}[h!]
\hspace{-1cm}
\resizebox{0.59\hsize}{!}{\includegraphics[angle=270]{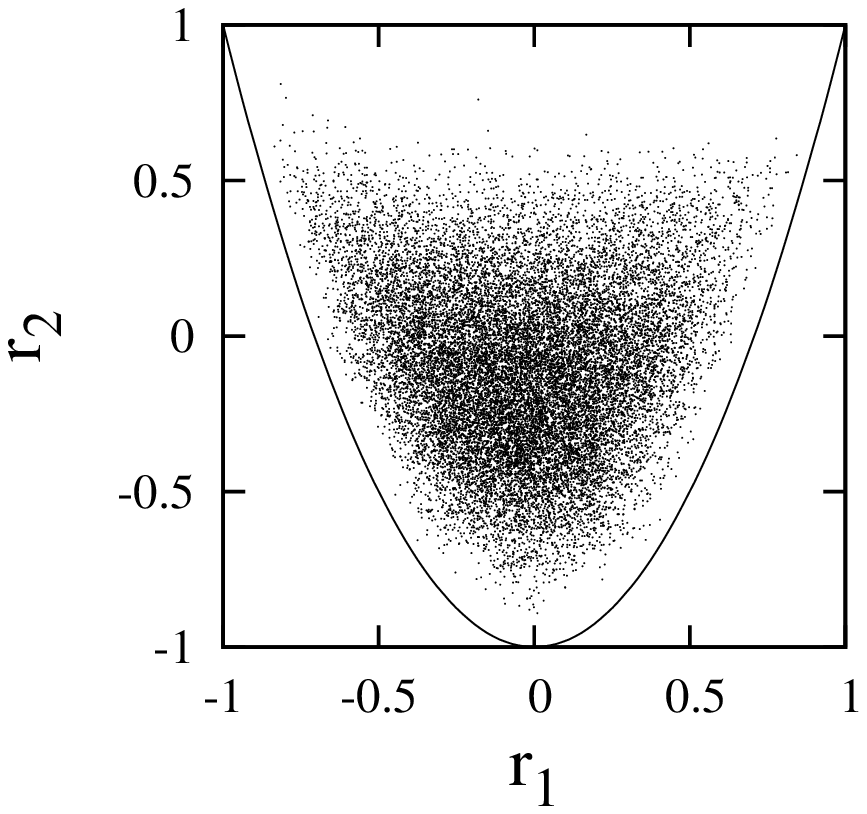}}\hspace{-2cm}
\resizebox{0.59\hsize}{!}{\includegraphics[angle=270]{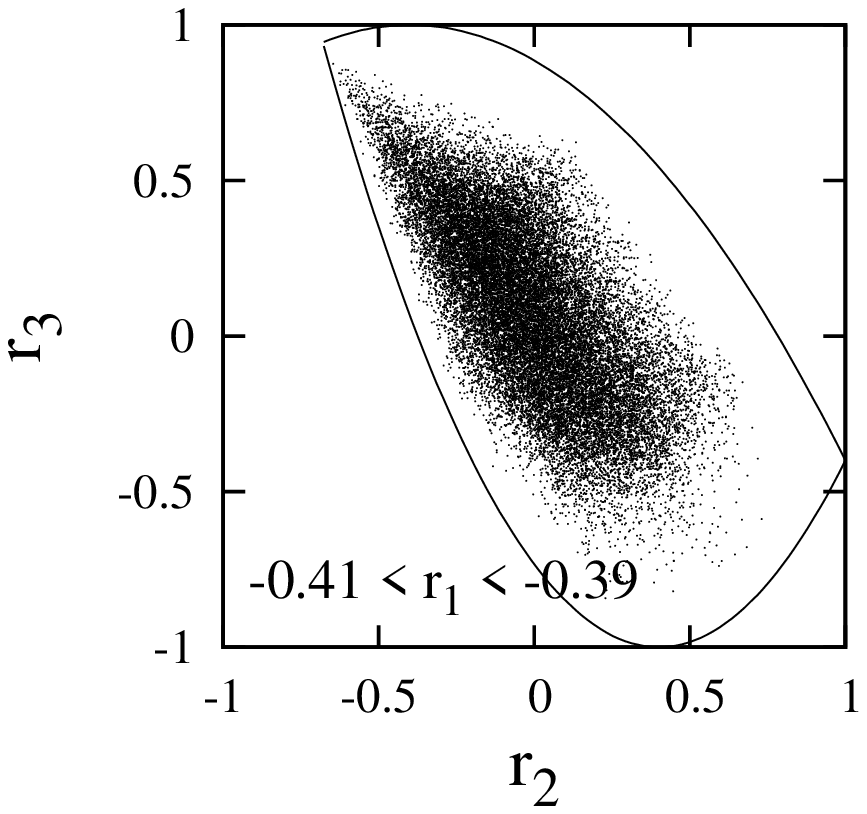}}
\caption{Constraints in the $r_1$-$r_2$-plane (left panel) and in the $r_2$-$r_3$-plane (left panel), where $r_1$ was constrained to \mbox{$-0.41< r_1 <-0.39$}. Each point corresponds to a correlation function measured from a realization of a one-dimensional Gaussian random field with a randomly drawn power spectrum and $N=16$ modes. Plotted as lines are the analytically determined constraints as given by Eqs.~(\ref{eq:constr2r}) and (\ref{eq:constr3r}).}
\label{fig:constraints}
\end{figure}

\begin{figure}[h!]
%\hspace{-1cm}
\resizebox{0.5\hsize}{!}{\includegraphics[angle=270]{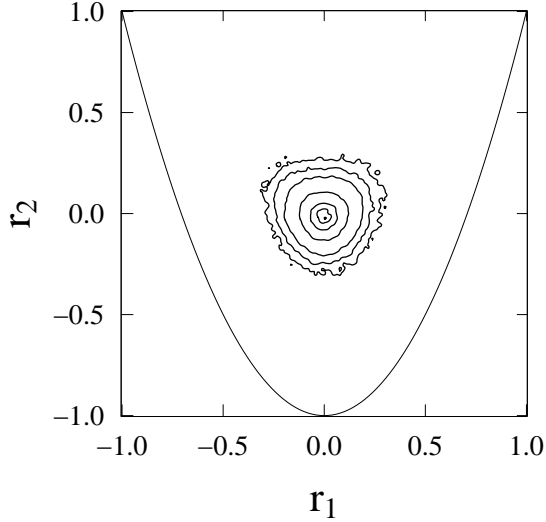}}\hspace{-2cm}
\caption{Distribution of ($r_1$, $r_2$) for correlation functions measured from simulated Gaussian random fields with $N=256$ and random power spectra (similar to Fig.~\ref{fig:constraints}). Even though the points lie well inside the admissible range, the shape of the distribution is similar to the shape of the allowed region.}
\label{fig:constraints2}
\end{figure}

We illustrate the constraints in the $r_1$-$r_2$-plane
(Eq.~\ref{eq:constr2r}) and in the $r_2$-$r_3$-plane for fixed $r_1$
(Eq.~\ref{eq:constr3r}) in Fig.~\ref{fig:constraints}. In order to
fully populate the allowed regions with data points, we do not use a
single power spectrum for all realizations of the random field, but
draw for each realization random positive values, uniformly distributed in $[0,1]$, for each $P_i$.  Note
that the data points do not completely fill out the regions allowed by
Eqs.~(\ref{eq:constr2r}) and (\ref{eq:constr3r}), but are constrained
to a slightly smaller region with piecewise linear boundaries. The
reason for this is that we have used a random field with $N=16$ modes,
whereas Eqs.~(\ref{eq:constr2r}) and (\ref{eq:constr3r}) have been
derived without reference to the specific way the random field is
created and are valid also for an infinite number of modes.

\begin{figure}
\resizebox{0.49\hsize}{!}{\includegraphics[angle=270]{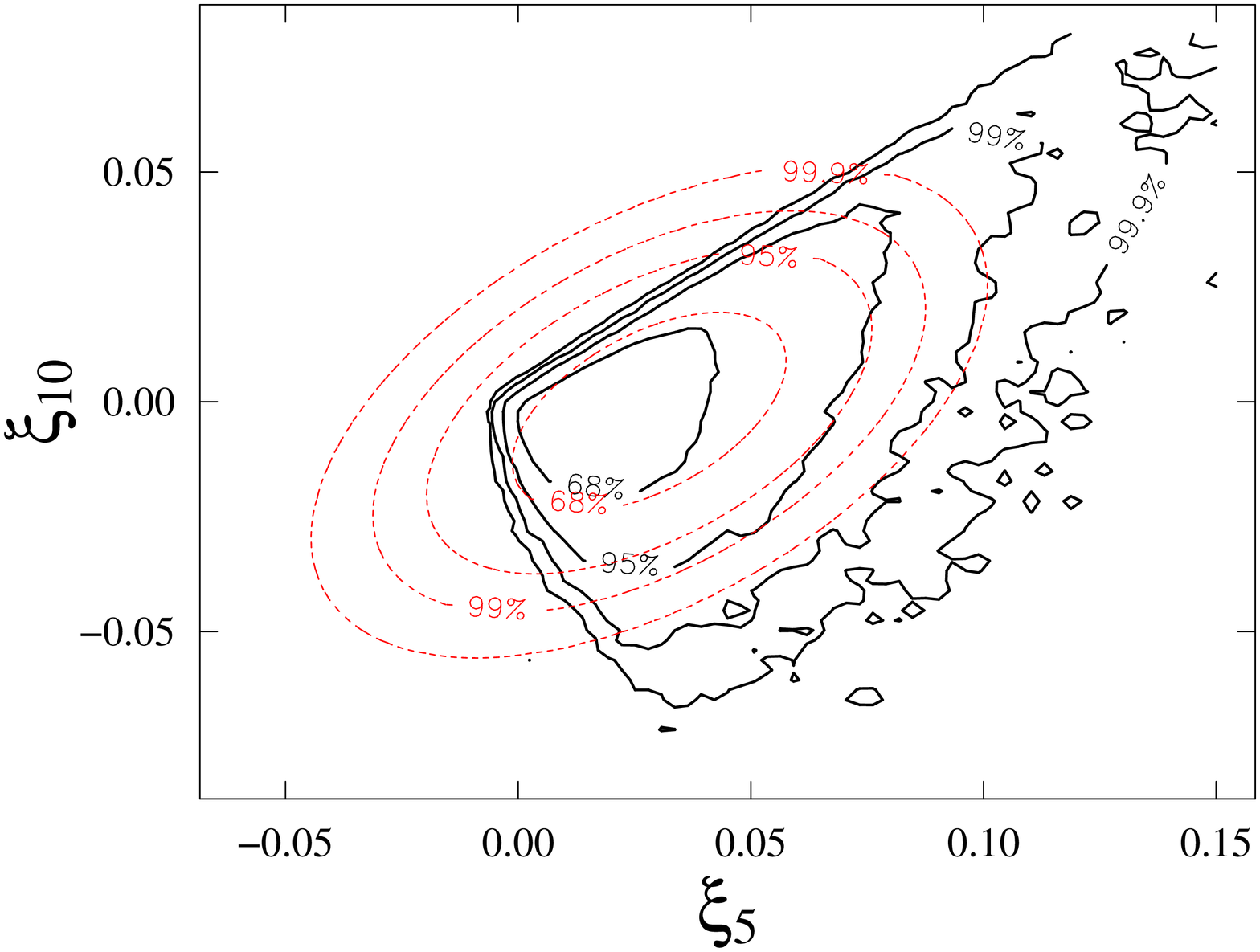}}\hfill
\resizebox{0.49\hsize}{!}{\includegraphics[angle=270]{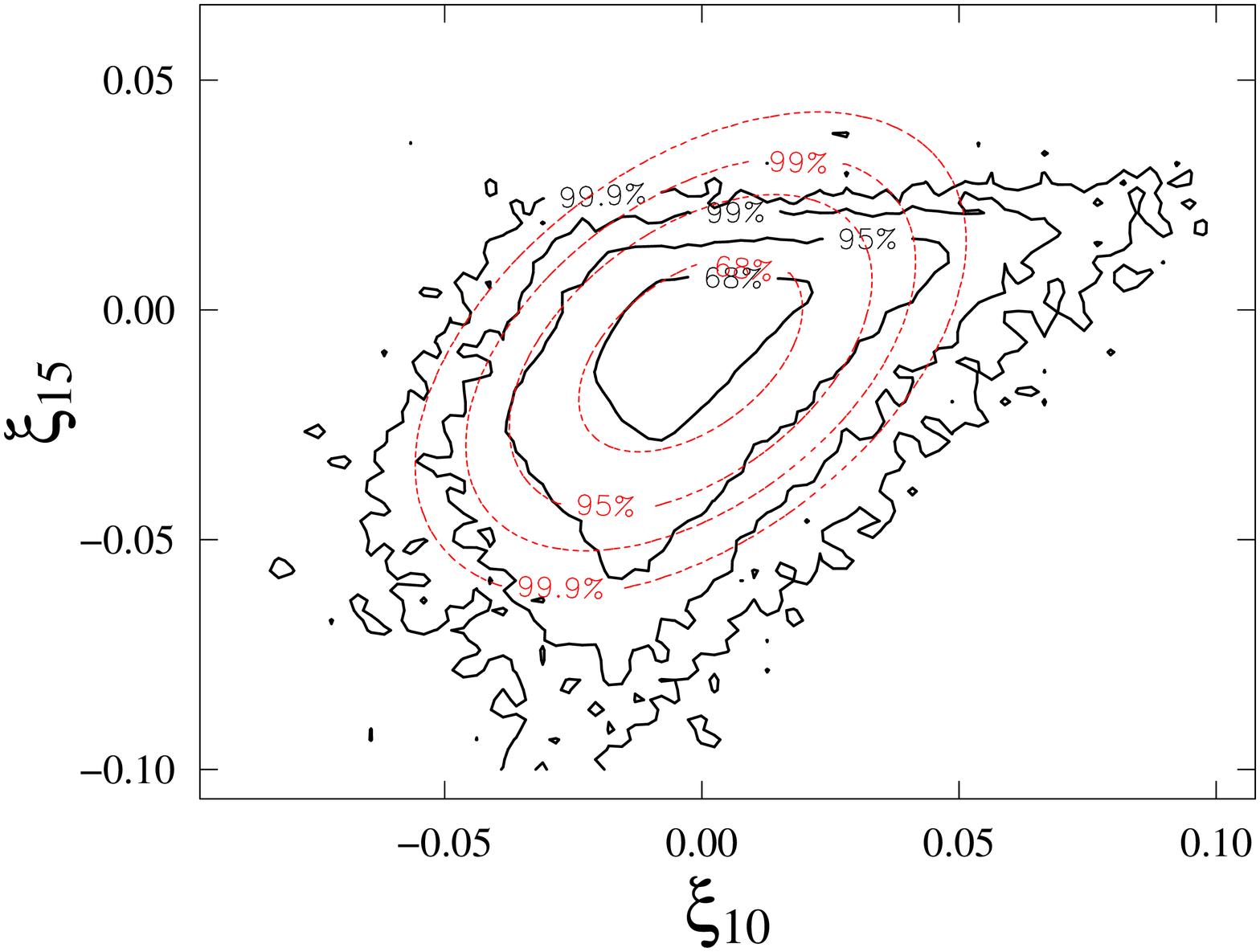}}
\caption{Marginalized likelihood of the correlation function components $\xi_{5}$ and $\xi_{10}$ (left panel) and $\xi_{10}$ and $\xi_{15}$ (right panel) as estimated from $10^5$ realizations of a one-dimensional Gaussian random field with $N=64$ modes and a Gaussian power spectrum (black contours). The corresponding Gaussian likelihood with covariance matrix as predicted from the power spectrum (Eq.~\ref{eq:cfcov}) is shown by the red dashed contours.}
\label{fig:xidist}
\end{figure}

The covariance matrix of the correlation function estimator given by
Eq.~(\ref{eq:cfest}) is given by
\be
	\tens{Cov}[\hat\xi]_{ij}=\frac{1}{N^2} \sum_{a=1}^N\sum_{b=1}^N\,\left( \xi_{|a-b|}\,\xi_{|a+i-b-j|} + \xi_{|a-b-j|}\,\xi_{|a-b+i|} \right) \elabel{cfcov}
\ee
With this, we can compare the true distribution of the correlation
function (as estimated from a large number of realizations of the
Gaussian field) to the commonly used Gaussian approximation to the
likelihood. As an example, we show the likelihoods
$\mathcal{L}(\xi_{5},\xi_{10})$ and $\mathcal{L}(\xi_{10},\xi_{15})$
in Fig.~\ref{fig:xidist}, where we have marginalized over all other
components of $\xi$. We have used a random field with $N=64$ modes and
a single Gaussian power spectrum.  Even though the estimated
likelihood and the Gaussian prediction have identical first and second
moments, a Gaussian likelihood clearly is a bad approximation to the
distribution of the correlation function.

\begin{figure}
\resizebox{0.95\hsize}{!}{\includegraphics[angle=270]{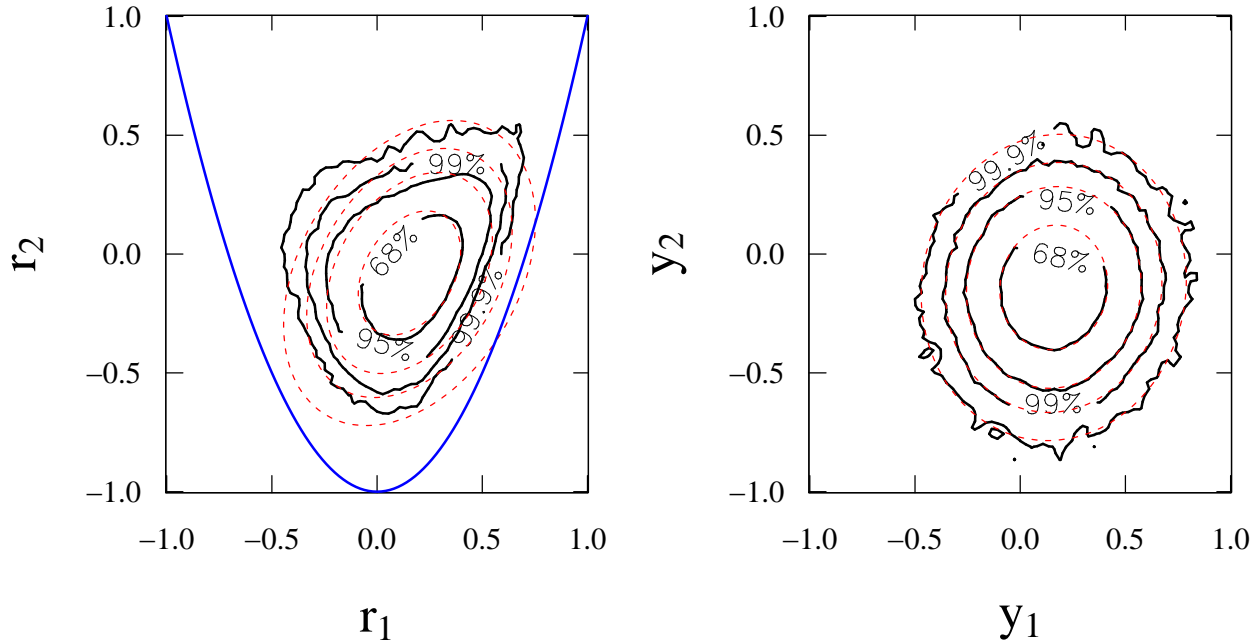}}\hfill
\caption{Marginalized distribution of $r_1$ and $r_2$ for a Gaussian random field with $N=64$ modes and Gaussian power spectrum (right panel) and the distribution of the corresponding transformed variables $y_1$ and $y_2$ (right panel). Shown as red dashed contours are the best-fitting bi-variate Gaussian distributions; in the left panel, the constraint given by Eq.~(\ref{eq:constr2r}) is shown as solid blue line.}
\label{fig:y_12}
\end{figure}

\begin{figure}
\resizebox{0.95\hsize}{!}{\includegraphics[angle=270]{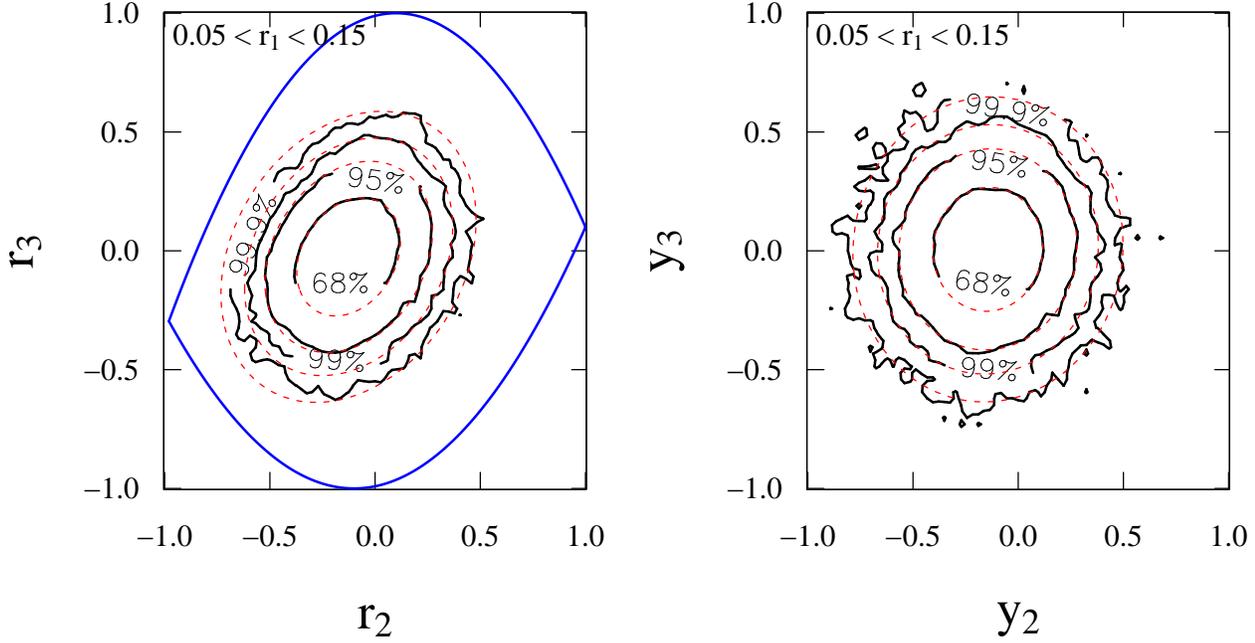}}\hfill
\caption{Distribution of $r_2$ and $r_3$ for a Gaussian random field
  with $N=64$ modes and Gaussian power spectrum (right panel), where
  we set $0.05<r_1<0.15$ and have marginalized over all other
  components of the correlation function, and the distribution of the
  corresponding transformed variables $y_2$ and $y_3$ (right
  panel). Shown as red dashed contours are the best-fitting bi-variate
  Gaussian distributions; in the left panel, the constraint given by
  Eq.~(\ref{eq:constr3r}) for $r_1=0.1$ is shown as solid blue line.}
\label{fig:y_23}
\end{figure}

We now investigate whether the transformation of variables from $r_n$
to $y_n$ given by Eqs.~(\ref{eq:rntoxn}) and (\ref{eq:xntoyn}) leads
to a likelihood of the $y_n$ that is better described by a
Gaussian. We find that this is indeed the case, as is illustrated in
Figs.~\ref{fig:y_12} and \ref{fig:y_23} for the marginalized
distributions in the 1-2-, and 2-3-planes. In the left panel, we
show the estimated distribution of the correlation coefficients,
whereas the right panel contains the distribution after the
transformation. The probability distribution in $r$-space `feels' the
presence of the boundary between allowed and forbidden regions, even
at the innermost contours which are quite well away from the boundary.
We also show the contours of the best-fitting bi-variate Gaussian for
comparison and see that the distribution in the $y$-space is much
better approximated by a Gaussian than the distribution of the
original correlation coefficients.  In addition, we note that the
transformation $r\mapsto y$ seems to reduce the correlations between
most of the components of the correlation function. It may be
suggested that the fact of the far more Gaussian distribution in the
transformed variable is related to the choice of the transformation
whose functional form is that of Fisher's z-transformation (Fisher
1915). 

Finally, we wish to assess the importance of the constraints in a more
realistic, two-dimensional setting and consider as an example a weak
lensing survey. Our strategy is as follows: we draw a large number of
realizations of the shear correlation function $\xi_+$, which is the
two-dimensional Fourier transform of the convergence power spectrum
$P_\kappa$ (Kaiser 1992; Bartelmann \& Schneider 2001) from a
multivariate Gaussian likelihood. The covariance matrix for the
likelihood function is computed under the assumption that the
convergence is a Gaussian random field using the methodology described
in Joachimi et al.~(2008). For each of these realizations, we
compute the matrix $\tens{A}$ (see Eq.\ts \ref{eq:Adef}).  We then use
this sample of correlation functions to compute a Monte-Carlo estimate
of the integral
\be
	\Lambda = \int {\rm d}^n \xi \; \mathcal{L}(\xi|p) \,H_{\rm pd}(\tens A)\;,	 \elabel{lensingtest}
\ee
where $\mathcal{L}(\xi|p)$ denotes the Gaussian likelihood as given by
Eq.~(\ref{eq:LLikeli}) and $H_{\rm pd}(\tens A)=1$ if $ \tens A$ is
positive semi-definite and $H_{\rm pd}(\tens A)=0$ otherwise. We test
$\tens A$ for positive semi-definiteness using the
Cholesky-decomposition (Press et al.\ 1992).  $\Lambda$ measures
the overlap of the Gaussian distribution with
the allowed region.  If $\Lambda < 1$, the Gaussian likelihood assigns
a finite probability to regions containing correlation functions that
do not correspond to a positive power spectrum and are therefore
forbidden by the constraints discussed in this paper. We can therefore
use $\Lambda$ as a rough indicator of the validity of the assumption
of a Gaussian likelihood. However, note that even if $\Lambda \approx
1$, the shape of the true likelihood might deviate significantly from
a Gaussian distribution.

\begin{figure*}
\vspace{1cm}
\sidecaption
\includegraphics[width=12cm]{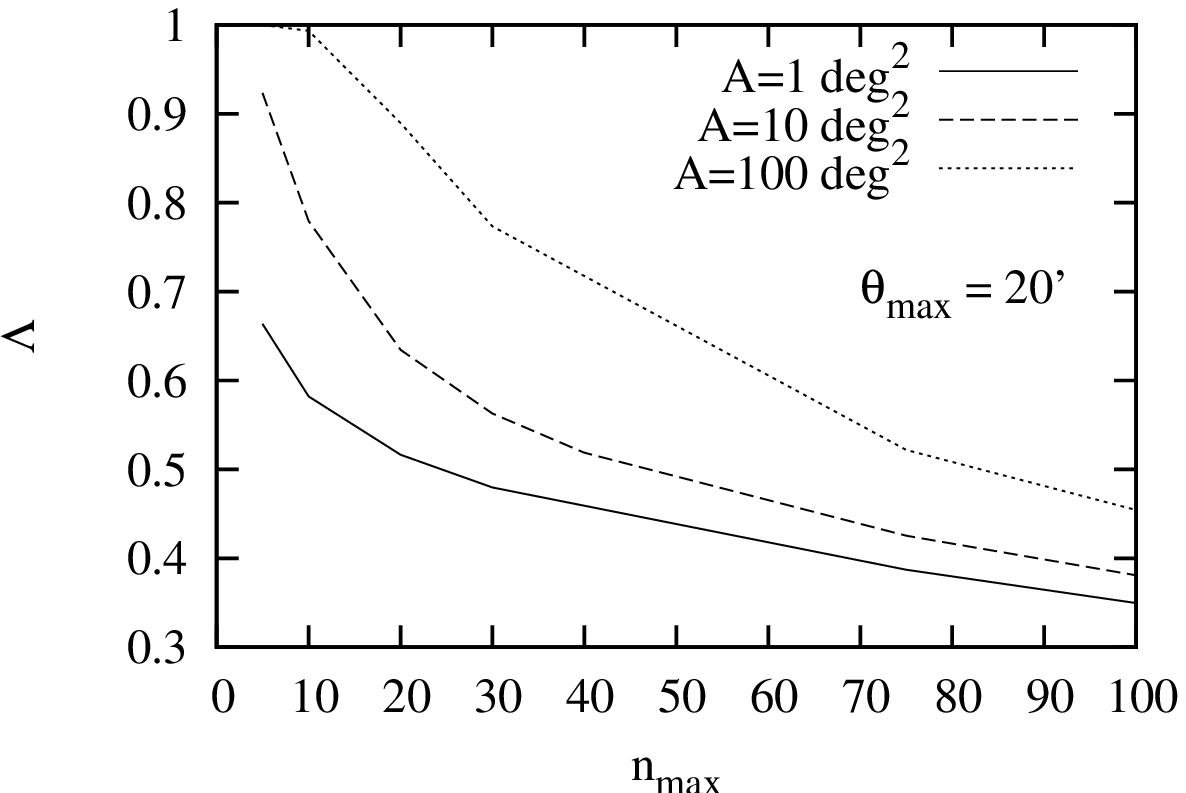}
\caption{Fraction $\Lambda$ (Eq.~\ref{eq:lensingtest}) of admissible
  (i.e.~positive semi-definite) shear correlation functions drawn from
  a Gaussian likelihood plotted versus the number of linear angular
  bins in the interval $0\leq\theta\leq20'$. Solid, dashed and dotted
  curves are for surveys with survey areas of 1, 10 and $100\,{\rm
    deg}^2$.\newline\newline\newline\newline} 
\label{fig:lambda}
\end{figure*}

For the numerical experiment, we choose a WMAP-5-like cosmology to
compute the shear correlation function and its covariance matrix
(approximating the shear field to be Gaussian), and
a redshift distribution of the source galaxies similar to the one
found for the CFHTLS-Wide (Benjamin et al.\ 2007). The results of
the numerical experiment are shown in Fig.~\ref{fig:lambda}, where we
plot $\Lambda $ as a function of the number of bins $n$ (linear
binning), keeping the maximum angular scale to which $\xi_+$ is
evaluated constant at $\theta_{\rm max} = 20'$. We display the results
for three different survey sizes ($1$, $10$ and $100\,{\rm
  deg}^2$). The constraints are more important for smaller survey
areas and a larger number of bins. This is because as the number of
bins increases, the constaints on the admissible values for the
following components of the correlation function become tighter, so
that the correlation function is basically determined by its first few
components. Increasing the noise (small area) or increasing the number
of bins therefore decreases the fraction of positive semi-definite
correlation functions that can be drawn from the Gaussian
likelihood. These results indicate that the likelihood of the
correlation function might deviate significantly from a Gaussian even
in realistic situations. It is expected that for real shear fields,
which are not Gaussian on the angular scales considered here, the
values of $\Lambda$ will deviate from unity even more.
We speculate that the non-Gaussianity found
in Hartlap et al.\ (2009) for the shear correlation functions
might at least partly be caused by the constraint of positive
semi-definiteness.

% \begin{figure}
% \resizebox{0.95\hsize}{!}{\includegraphics[angle=270]{figs/N32_gauss_rescaledCF_1_6.ps}}\hfill
% \caption{Marginalised distribution of $r_1$ and $r_6$ for a Gaussian random field with $N=64$ modes and Gaussian power spectrum (right panel), and the distribution of the corresponding transformed variables $y_1$ and $y_6$ (right panel). Shown as red dashed contours are the best-fitting bi-variate Gaussian distributions.}
% \label{fig:y_16}
% \end{figure}

\section{Conclusions and outlook}
We have considered constraints on correlation functions that need to
be satisfied in order for the correlation function to correspond to a
non-negative power spectrum. Using the covariance matrix method, we
have derived explicit expressions for the upper and lower bounds on
the correlation coefficients $r_n$; these were derived for the case
that the spatial sampling of $\xi$ occurs at points $x_j= j x$. This
method yields optimal constraints for the correlation of
one-dimensional random fields, whereas they are not optimal for
higher-dimensional homogeneous and isotropic random fields. We have
indicated a method with which such optimal constraints can in
principle be derived for higher-dimensional fields and for non-linear
spacing of grid points, and presented a few simple applications of
this method; however, up to now we have not obtained a systematic
method how to derive explicit upper and lower bounds on the $r_n$ in
these cases. Finding those will be of considerable interest since they
are expected to be tighter than the corresponding bounds derived for
the 1-D case.

Using cosmic shear as an example, we have demonstrated that the
Gaussian probability ellipsoid, which is obtained under the assumption
that the likelihood function of the correlation function is given by a
Gaussian characterized by the covariance matrix, significantly spills
over to the forbidden region of correlation functions. This effect is
even more serious than considered here, since we have used for this
analysis the constraints from Sect.\ts\ref{sc:2} which are not
optimal, as shown in Sect.\ts\ref{sc:4}.  Hence, from this argument
alone we conclude that the assumption of a Gaussian likelihood is not
very realistic and probably lead to erroneous estimates of parameters
and their confidence regions.

Even if the Gaussian likelihood ellipsoid is contained inside the
allowed region, the true likelihood deviates from a
Gaussian, as simple numerical experiments with one-dimensional Gaussian
random fields have shown. Surprisingly, the shape of the resulting likelihood
contours  have some resemblance to the shape of the boundary of the
allowed region even if the size of the probability distribution is
considerably smaller than the allowed region. The origin of this
result is not understood. Most likely, the likelihood of the
correlation function depends not only on its covariance matrix, but
also on higher-order moments. 

A non-linear coupled transformation of the correlation coefficients
leads to a distribution that appears much more Gaussian (in the
transformed variables), and there may be a connection of this fact to
the Independent Components analyzed in Hartlap et al.\ (2009). Indeed,
in these transformed variables, the likelihood not only is much more
Gaussian, but also less correlated, which supports the hypothesis
about a connection between the constraints derived here and the ICA
analysis of Hartlap et al.\ (2009). More extensive numerical tests may
yield better insight into this connection. It must be stressed that
such a result, if it can be obtained, would be of great importance,
given that the determination of multi-variate probability
distributions from numerical simulations is prohibitively expensive.

An alternative route for understanding the connection between the
constraints derived here and the shape of the likelihood function is
the explicit calculation of the multivariate probability distribution
of the correlation function $\xi(x_j)$ for a Gaussian random
field. The constraints on the $r_j$ should be explicitly present in
this probability distribution. Work on these issues is currently in
progress.

The results of this paper can most likely be generalized to random
fields which are not scalars, e.g., the polarization of radiation, or
the orientation of objects. The cosmic shear correlation function
(e.g., Kaiser 1992; Schneider et al.\ 2002)
$\xi_+$ which has been considered in Sect.\ts\ref{sc:6} is equivalent
to the correlation function of the underlying surface mass density
$\kappa$, and thus the correlation behaves in the same way as that of
a scalar field. However, the other cosmic shear correlation function
$\xi_-$ is qualitatively different, being a spin-4 quantity, and for
which the filter function in (\ref{eq:xiofP-2D}) is replaced by ${\rm
  J}_4(kx)$. 

The foremost aim of this paper was the derivation of exact constraints
on correlation functions; in contrast, we have not considered methods
for measuring a correlation function from data. For example, in many
cases the correlation function cannot be measured at zero lag, so that
the correlation coeffcients $r=\xi(x)/\xi(0)$ can then not be
determined directly. Furthermore, one derives the correlation function
from data in a given volume, and thus the measured correlation
function will deviate from the ensemble average, even in the absence
of noise. This effect has two different aspects: suppose for a moment
that the observed field is one-dimensional and forced (or assumed) to
be periodic. If the correlation function on such a field is measured
using the definition (\ref{eq:cfest}), then the measured correlation
function will deviate from the ensemble average, but it will still
correspond to a non-negative power spectrum, given by the square of
the Fourier transform of the realization of the field. Hence, in such
a case, every measured correlation function will satisfy the
constraints derived in Sect.\ts\ref{sc:2}. If the field has more than
one dimension, but is still periodic, the correlation function
measured by a generalization of (\ref{eq:cfest}) to higher dimensions
will still obey the constraints from Sect.\ts\ref{sc:2}, for the same
reason. However, the power spectrum of the realization of the field
will in general not be isotropic, and thus the considerations of
Sect.\ts\ref{sc:4} do not apply strictly. In the more realistic case
where periodicity cannot be assumed, one cannot measure the
correlation function by `wrapping around' as in (\ref{eq:cfest}); in
this case, there are `boundary effects', which are the stronger the
more the separation approaches the size of the data field. Then, there
exists not necessarily a non-negative power spectrum related to the
measured correlation function through (\ref{eq:xiofP}), and the
constraints may not apply strictly. How important these effect are
needs to be studied with a more extended set of simulations.

%As a side remark, the matrix $\tens A$ (see Eq.\ts\ref{eq:Adef}) is a
%symmetric Toeplitz matrix, and the bounds obtained for the
%correlation coefficients $r_j$ are thus the conditions for the
%positive definiteness of such matrices. 

\section*{Acknowledgements}
We thank Martin Kilbinger and Cristiano Porciani for constructive
comments on the manuscript. This work was supported by the Deutsche
Forschungsgemeinschaft in the framework of the Priority Programme 1177
on `Galaxy Evolution' and of the Transregional Cooperative Research
Centre TRR 33 `The Dark Universe', and by the Bonn-Cologne Graduate
School of Physics and Astronomy.
%\begin{appendix}
%\section{\label{Sc:A}The matrix $C$}
%\end{appendix}

%/////////////////////////////////////////////////////////////////////

\end{document}